\documentclass[showpacs,preprintnumbers,amsmath,amssymb,floatfix]{revtex4}

\usepackage{epsf}
\usepackage[hang,nooneline]{subfigure}

\usepackage{graphicx}

\newcommand{\insertplot}[5]{\begin{figure}
 \hfill\hbox to 0.05in{\vbox to #5in{\vfill
 \inputplot{#1}{#4}{#5}}\hfill}
 \hfill\vspace{-.1in}
 \caption{#2}\label{#3}
 \end{figure}}
 \newcommand{\inputplot}[3]{
 \special{ps: plotfile #1}
}
\newcounter{fig}   

\newcommand{\vphi}{\varphi}
\newcommand{\vepsilon}{\varepsilon}

\usepackage{graphicx}

\begin{document}

\title{
Radially excited rotating black holes in Einstein-Maxwell-Chern-Simons theory
}
 \vspace{1.5truecm}
\author{
{\bf Jose Luis Bl\'azquez-Salcedo$^1$},
{\bf Jutta Kunz$^1$},\\
{\bf Francisco Navarro-L\'erida$^2$},
{\bf Eugen Radu$^3$}\\
\vspace{0.5truecm}
$^1$
 Institut f\"ur  Physik, Universit\"at Oldenburg\\ Postfach 2503,
D-26111 Oldenburg, Germany\\
$^2$
Dept.~de F\'{\i}sica At\'omica, Molecular y Nuclear, Ciencias F\'{\i}sicas\\
Universidad Complutense de Madrid, E-28040 Madrid, Spain\\
$^3$
Departamento de F\'\i sica da Universidade de Aveiro and I3N\\
Campus de Santiago, 3810-183 Aveiro, Portugal
}
\vspace{0.5truecm}

\vspace{0.5truecm}

\vspace{0.5truecm}
\date{
\today}

\begin{abstract}
Rotating black holes in Einstein-Maxwell-Chern-Simons theory 
possess remarkable features, when the Chern-Simons coupling constant
reaches a critical value. Representing single asymptotically flat 
black holes with horizons of spherical topology, 
they exhibit non-uniqueness. 
In particular, there even exist extremal and non-extremal
black holes with the same sets of global charges.
Both extremal and non-extremal black holes
form sequences of radially excited solutions, that can be
labeled by the node number of the magnetic gauge potential function.
The extremal Reissner-Nordstr\"om solution is no longer always located on
the boundary of the domain of existence of these black holes,
and it neither remains the single extremal solution with vanishing 
angular momentum.  Instead a whole sequence of rotating extremal $J=0$ 
solutions is present, whose mass converges towards the mass of the 
Reissner-Nordstr\"om solution. These radially excited extremal solutions 
are all associated with the same near horizon solution.  Moreover, 
there are near horizon solutions that are not realized as global solutions.
\end{abstract}

\maketitle

  \medskip


\section{Introduction}

In four spacetime dimensions vacuum black holes
and black holes in the presence of an electomagnetic field
possess very remarkable properties, as formulated
by a number of important theorems.
The black hole uniqueness theorem, for instance,
states that asymptotically flat non-degenerate electrovac black holes are 
uniquely described by their global charges: the mass,
the angular momentum and the electric (and magnetic) charge
\cite{Israel:1967za,Robinson:1975bv,Mazur:1982db,Chrusciel:2012jk}.

When going to higher dimensions, the Schwarzschild solutions
and the Reissner-Nordstr\"om (RN) solutions possess simple generalizations,
while the Kerr solutions are generalized by the family of
Myers-Perry (MP) solutions \cite{Myers:1986un,Myers:2011yc}.
In $D$ dimensions, these solutions possess $N=\lfloor (D-1)/2 \rfloor$
independent spatial planes and thus also 
$N$ independent angular momenta.
In five dimensions both angular momenta are bounded for a given mass,
a feature which does not hold for $D>5$. 

The higher-dimensional generalizations of the Kerr-Newman solutions
are not known in closed form. However, they have been studied perturbatively
\cite{Aliev:2004ec,Aliev:2005npa,Aliev:2006yk,NavarroLerida:2007ez,Aliev:2008bh,Sheykhi:2008bs,Allahverdizadeh:2010xx,Allahverdizadeh:2010fn}
and numerically 
\cite{Kunz:2005nm,Kunz:2006eh,Blazquez-Salcedo:2013yba,Blazquez-Salcedo:2013wka}.
In odd dimensions, when all angular momenta possess equal magnitude,
the angular coordinates factorize, leading to substantial
simplifications of the perturbative and numerical studies.
Moreover, for extremal black holes the near-horizon solutions have been 
investigated \cite{Kunduri:2013gce,Blazquez-Salcedo:2013wka}.
In the case of black hole solutions with equal magnitude
angular momenta in odd dimensions,
there are two branches of near-horizon solutions,
the MP branch and the RN branch, intersecting at a critical point. 
Interestingly, for both branches, only a part of the near-horizon solutions 
also correspond to global solutions, thus not all
Einstein-Maxwell (EM) near-horizon solutions have global counterparts.

In odd dimensions, a Chern-Simons (CS) term can be added to the
action, breaking the charge reversal symmetry for even $N=(D-1)/2$.
The resulting Einstein-Maxwell-Chern-Simons
(EMCS) theories possess intriguing sets of black hole solutions
\cite{Breckenridge:1996is,Cvetic:2004hs,Chong:2005hr,Kunz:2005ei,Kunz:2006yp,Blazquez-Salcedo:2013muz}.
In five dimensions for the special case of the CS coupling constant
$\lambda=\lambda_{\rm SG}$, as obtained for minimal supergravity,
the general set of charged rotating black hole solutions
is known in closed form \cite{Chong:2005hr}.

Carrying opposite angular momenta,
the BMPV \cite{Breckenridge:1996is} black holes represent 
a subset of these solutions.
Emerging from one of the charge symmetric extremal RN solutions,
they form a branch of stationary extremal black hole solutions
with vanishing horizon angular velocities.
The magnitude of their (equal magnitude) angular momenta can be increased, 
while keeping their mass and charge fixed, until a singular solution
with vanishing area is encountered.
Recall, that in four dimensions stationary EM black holes 
with a non-rotating horizon are static. 
It is the presence of the CS term, which allows for 
non-static black holes with vanishing horizon angular momenta.
Here the frame dragging effects at the inside and outside of the horizon
precisely cancel at the horizon in the case of the BMPV black holes
\cite{Gauntlett:1998fz}.

When the CS coupling constant $\lambda$ is increased beyond its supergravity
value, the set of black hole solutions acquires interesting new features.
First, for $\lambda_{\rm SG} \le \lambda < 2 \lambda_{\rm SG}$,
counterrotation sets in, $i.e.$, within a certain region of
the domain of existence of the solutions, the horizon angular velocity
and the angular momentum have opposite signs \cite{Kunz:2005ei,Kunz:2006yp}.
As a consequence, extremal static black holes can become unstable
with respect to rotation. Indeed, for fixed electric charge
the mass can decrease with increasing magnitude of the angular momentum.
Thus supersymmetry marks the borderline between stability and instability
for EMCS black holes
\cite{Gauntlett:1998fz,Kunz:2005ei}.

Next, for $\lambda > 2 \lambda_{\rm SG}$, the extremal RN solution
remains part of the boundary of the domain of existence only for 
one sign of the charge.
For the other sign of the charge, the extremal RN solution
resides inside the boundary of the domain of existence.
As boundary solution with vanishing angular momentum the static RN solution
is now replaced by a set of two stationary solutions, whose global
angular momenta vanish \cite{Kunz:2005ei}.
Thus EMCS theory allows for black holes with rotating horizon
but vanishing angular momenta, where the contributions 
to the global angular momenta precisely balance.

These two $J=0$ rotating extremal solutions form only the lowest mass
solutions of a whole sequence of excited $J=0$ rotating extremal solutions,
which possess an increasing number of radial nodes in one of the metric 
and one of the gauge field functions \cite{Blazquez-Salcedo:2013muz}.
For a fixed value of the charge, the mass of this
sequence of solutions converges towards the mass of the 
corresponding extremal RN black hole.
Since these excited extremal black hole solutions are located inside the domain
of existence these solutions represent a new type of violation of uniqueness:
there are extremal and non-extremal black holes with the same sets of global
charges \cite{Blazquez-Salcedo:2013muz}.
The violation of uniqueness among non-extremal black holes is of course
also present \cite{Kunz:2005ei}.

When the full sets of extremal solutions are considered, an intricate web
of branches arises. These branches of global black hole solutions can be compared with
the branches of near-horizon solutions. 
The structure of the latter is much simpler.
Comparison of the two sets shows, that there are near-horizon solutions,
which correspond to (i) no global solutions, (ii) exactly one global solution,
(iii) more that one global solution
\cite{Blazquez-Salcedo:2013muz}.

Here we give a detailed account of the 
astounding properties of EMCS black hole solutions
in five dimensions, treating the CS coupling constant as a free parameter,
 a task which, to our knowledge, has not been yet considered in the literature.
 Special attention will be paid to extremal solutions, although some properties of
non-extremal configurations will be discussed as well. 
The paper is organized as follows:
in Section II we present the action, the Ans\"atze and the charges.
We discuss the near-horizon solutions in Section III,
where the near-horizon formalism must be employed with care 
because of the presence of the CS term.
We present our numerical procedure for the global solutions in Section IV,
discuss the boundary conditions and provide expansions for the functions.
Our numerical results together with a comparison of global and near-horizon
solutions are given in Section V. We end with a brief conclusion
and outlook in Section VI.

\section{Action, Ans\"atze and Charges}

We focus our study on black holes in Einstein-Maxwell-Chern-Simons (EMCS) theory.
Here we briefly review the action and the general set of equations of motion.
We then present the appropriate Ans\"atze to obtain rotating black holes with equal magnitude
angular momenta.
Next we recall the general formulae for the global charges and the horizon properties
of these black holes, as well as their scaling symmetry.

\subsection{Einstein-Maxwell-Chern-Simons action}

The action of Einstein-Maxwell-Chern-Simons theory in five dimensions reads
\begin{equation} \label{EMCSac}
I= \frac{1}{16\pi G_5} \int d^5x\biggl[ 
\sqrt{-g} \, (R -
\frac{1}{4}F_{\mu \nu} F^{\mu \nu}
)
-
\frac{\lambda}{12\sqrt{3}}\,\varepsilon^{\mu\nu\alpha\beta\gamma}A_{\mu}F_{\nu\alpha}F_{\beta\gamma} 
 \biggr ]  , 
\end{equation}
where $R$ is the curvature scalar,
$G_5$ is Newton's constant in five dimensions,
$A_\mu $ is the gauge potential with field strength tensor $ F_{\mu \nu}
= \partial_\mu A_\nu -\partial_\nu A_\mu $,
and ${ \lambda}$ is the Chern-Simons (CS) coupling constant. 
In the following we employ units such that ${16\pi} \, G_5=1$.
Note, that for $\lambda=\lambda_{\rm SG}=1$ the action corresponds to the 
bosonic sector of minimal supergravity.
For $\lambda=\lambda_{\rm EM}=0$ the action corresponds to
Einstein-Maxwell theory.

Variation of the action with respect to the metric leads to the 
Einstein equations 
\begin{equation}
\label{Einstein_equation}
G_{\mu\nu}=\frac{1}{2} T_{\mu\nu},
\end{equation}
where the stress-energy tensor is given by
\begin{equation}
T_{\mu\nu} = F_{\mu\rho} {F_\nu}^\rho 
  - \frac{1}{4} g_{\mu \nu} F_{\rho \sigma} F^{\rho \sigma}.
\end{equation}
Variation with respect to the gauge potential leads to the 
Maxwell-Chern-Simons equations
\begin{equation}
\label{Maxwell_equation}
\nabla_{\nu} F^{\mu\nu} + \frac{\lambda}{4\sqrt{3}}\varepsilon^{\mu\nu\alpha\beta\gamma}F_{\nu\alpha}F_{\beta\gamma}=0.
\end{equation}

\subsection{Ans\"atze}

We consider stationary black holes,
which represent generalizations of the five-dimensional
Myers-Perry solutions \cite{Myers:1986un} to EMCS theory. 
Hence these black holes possess a spherical horizon topology.
Their Killing vectors are
\begin{equation}
\xi \equiv \partial_t , \ \ \
\eta_{(1)} \equiv \partial_{\varphi_1} , \ \ \
\eta_{(2)} \equiv \partial_{\varphi_2} ,
\label{killing} 
\end{equation}
 with $t$ the time coordinate and $\varphi_{1,2}$ angular directions.
With each azimuthal symmetry an angular momentum $J_{(k)}$, $k=1,2$, is associated.
In general, both angular momenta are independent.
 
Here we restrict to the case where both angular momenta have equal magnitude,
$|J_{(1)}|=|J_{(2)}|=|J|$. The spacetime then becomes a
cohomogeneity-1 manifold,
 with an enhancement of the isometry group  from $R_t\times U(1)^2$ to $R_t\times U(2)$. 
 Hence the angular dependence of the metric and the
gauge potential can be explicitly given. 
Then the metric may be parametrized by the Ansatz 
\begin{eqnarray}
\label{metric0}
&&ds^2 = -F_0(r) dt^2 + F_1(r)dr^2 + F_2(r) d\theta^2   + F_3(r) \sin^2\theta \left( \varepsilon_1 d \varphi_1 -W(r) dt
\right)^2 \nonumber 
\\  && 
+ F_3(r) \cos^2\theta \left(\varepsilon_2 d \varphi_2
  -W(r)dt \right)^2 +(F_2(r)-F_3(r) ) \sin^2\theta \cos^2\theta(\varepsilon_1 d \varphi_1  -\varepsilon_2 d \varphi_2)^2,
\end{eqnarray}
where $\theta \in [0,\pi/2]$, $\varphi_1 \in [0,2\pi]$ and $\varphi_2 \in [0,2\pi]$,
and $\vepsilon_k = \pm 1$ denotes the sense of rotation
in the $k$-th orthogonal plane of rotation.
Note also that the line element Eq. (\ref{metric0}) still possesses a residual metric gauge freedom.

The corresponding Ansatz for the gauge potential is given by 
\begin{equation}
A_\mu dx^\mu  = a_0(r) dt + a_{\varphi}(r) (\sin^2 \theta \varepsilon_1 d\varphi_1+\cos^2 \theta \varepsilon_2 d\varphi_2).
\end{equation}

\subsection{Known solutions and parametrization used in the numerics}

 For the general family of EMCS solutions within the above Ans\"atze, only the following special cases
  are known in closed form: 

$(i)$ Vacuum rotating black holes (Myers-Perry solution). This solution is
valid for every value of $\lambda$, since the Maxwell and the Chern-Simons terms 
vanish identically.
The functions which enter the line element Eq. (\ref{metric0}) are
\begin{eqnarray}  
 &&
 F_0(r)=\frac{1-\frac{M}{6\pi^2 r^2}+\frac{3J^2}{8\pi^2 M r^4}}{1+ \frac{3J^2}{8\pi^2 M r^4}},~~
 F_1(r)=\frac{1}{1+ \frac{3J^2}{8\pi^2 M r^4}},~~F_2(r)=r^2,
 \\
 \nonumber
 &&
 F_3(r)=r^2(\frac{3J^2}{8\pi^2 M r^4}),~~W(r)=\frac{J}{4\pi^2 r^4(1+ \frac{3J^2}{8\pi^2 M r^4})},
\end{eqnarray} 
while $a_{\varphi}(r)=a_0(r)=0$.
Also, $M$ and $J$ are the mass and the angular momentum of the solutions.

$(ii)$ Charged static black holes (Reissner-Nordstr\"om solution). Again this
solution is valid for every value of $\lambda$. In the static
case, the magnetic component of the gauge field vanishes,  $a_{\varphi}(r)=0$,
and the Chern-Simons term does not contribute.
The expressions of the other functions which enter the solution are 
\begin{eqnarray} 
F_0(r)=\frac{1}{F_1(r)}=1-\frac{M}{6\pi^2 r^2}+\frac{Q^2}{48\pi^2 r^4},
~~F_2(r)=F_3(r)=r^2,~~W(r)=0,~~{\rm and}~~a_0(r)=\frac{Q}{4\pi^2r^2},
\end{eqnarray} 
with $Q$ the electric charge.

$(iii)$ The general set of charged rotating solutions for $\lambda=\lambda_{\rm SG}=1$
\cite{Chong:2005hr} with the extremal BMPV black holes \cite{Breckenridge:1996is}
as a special case.
These black holes have
\begin{eqnarray} 
&&
F_0(r)= \frac{1-\frac{2(p-q)}{r^2}+\frac{2j^2p+q^2}{r^4}}{1+\frac{2j^2p}{r^4}-\frac{ j^2 q^2}{r^6}},~~
F_1(r)= \frac{1}{1-\frac{2(p-q)}{r^2}+\frac{2j^2p+q^2}{r^4}},~~F_2(r)=r^2,
\\
\nonumber
&&
F_3(r)= (1+\frac{2j^2p}{r^4}-\frac{ j^2 q^2}{r^6})r^2,~ 
W(r)=\frac{2p-q-\frac{q^2}{r^2}}{1+\frac{2j^2p}{r^4}-\frac{ j^2 q^2}{r^6}}\frac{j}{r^4},
~{\rm and}~a_{\varphi}(r)=-\frac{\sqrt{3} jq}{r^2},~a_0(r)=\frac{\sqrt{3}q}{r^2},
\end{eqnarray} 
with $p=\frac{M+\sqrt{3}Q}{12\pi^2}$, 
$j=\frac{3J}{2M+\sqrt{3}Q}$,
$q=\frac{Q}{4\sqrt{3}\pi^2}$.

In the numerics, we have found it useful to 
use quasi-isotropic coordinates by choosing a metric gauge with $F_2(r)=F_1(r)r^2$
and take 
$F_0(r)=f(r),$
$F_1(r)=\frac{m(r)}{f(r)}$,
$F_3(r)=\frac{n(r)}{f(r)}r^2$,
$W(r)=\frac{w(r)}{r}$.
This results in the following expression of the line element in terms of
four unknown functions
\begin{eqnarray}
\label{metric}
&&ds^2 = -f(r) dt^2 + \frac{m(r)}{f(r)}(dr^2 + r^2 d\theta^2)  + \frac{n(r)}{f(r)}r^2 \sin^2\theta \left( \varepsilon_1 d \varphi_1 -\frac{\omega(r)}{r}dt
\right)^2 \nonumber \\  && + \frac{n(r)}{f(r)}r^2 \cos^2\theta \left(\varepsilon_2 d \varphi_2
  -\frac{\omega(r)}{r}dt \right)^2 + \frac{m(r)-n(r)}{f(r)}r^2 \sin^2\theta \cos^2\theta(\varepsilon_1 d \varphi_1  -\varepsilon_2 d \varphi_2)^2,
\end{eqnarray}
 originally proposed in 
 \cite{Kunz:2005nm,Kunz:2006eh}.

\subsection{Global charges}

We here consider stationary asymptotically flat black holes.
Their total mass $M$ and angular momentum $J_{(k)}$ can be obtained
directly from the Komar expressions associated with the Killing vector fields
\begin{equation}
M = -  \frac{3}{2} \int_{S_{\infty}^{3}} \alpha  
\ , \label{Kmass} \end{equation}
\begin{equation}
J_{(k)} =   \int_{S_{\infty}^{3}} \beta_{(k)} 
\ , \label{Kang} \end{equation}
where $\alpha_{\mu_1 \mu_2 \mu_3} \equiv \epsilon_{\mu_1 \mu_2 \mu_3
  \rho \sigma} \nabla^\rho \xi^\sigma$,
and
$\beta_{ (k) \mu_1 \mu_2 \mu_3} \equiv \epsilon_{\mu_1 \mu_2 \mu_3
  \rho \sigma} \nabla^\rho \eta_{(k)}^\sigma$.
For equal-magnitude angular momenta $J_{(k)}=\varepsilon_k J$, $k=1, 2$.

The electric charge $Q$ is given by
\begin{equation}
Q= - \frac{1}{2} \int_{S_{\infty}^{3}} \left( \tilde F 
+\frac{\lambda}{\sqrt{3}} A \wedge F \right)
= - \frac{1}{2} \int_{S_{\infty}^{3}} \tilde F 
\ , \label{charge} \end{equation}
where
${\tilde F}_{\mu_1 \mu_2 \mu_3} \equiv  
  \epsilon_{\mu_1 \mu_2 \mu_3 \rho \sigma} F^{\rho \sigma}$.
The magnetic moment $\mu_{\rm mag}$ is determined from the
 asymptotic expansion of the gauge potential $a_{\varphi}(r)$, see Section IV B.
The gyromagnetic ratio $g$ is then obtained from the
magnetic moment $\mu_{\rm mag}$ via
\begin{equation}
{\mu_{\rm mag}}=g \frac{Q J}{2M}
\ . \label{gyro} 
\end{equation}

\subsection{Horizon Properties}

The black hole horizon is located at $r=r_{\rm H}$ 
and rotates with angular velocity $\Omega_{\rm H}$.
It is a Killing horizon, where
the Killing vector $\zeta = \partial_t + \Omega_{\rm H} (
\varepsilon_1 \partial_{\varphi_1} +
\varepsilon_2 \partial_{\varphi_2})$ becomes null
and orthogonal to the other Killing vectors
\begin{eqnarray}
(\zeta^2)|_{\cal H}=0, \ \ \
(\zeta\cdot\partial_t)|_{\cal H}=0,\ \ \
(\zeta\cdot\partial_{\varphi_1})|_{\cal H}=0, \ \ \
(\zeta\cdot\partial_{\varphi_2})|_{\cal H}=0.
\end{eqnarray}
Note, that these expressions impose conditions on the metric functions.

In particular,
for the parametrization Eq. (\ref{metric}) employed in the numerics,
 the metric function $f(r)$ must vanish at the horizon.
\begin{eqnarray}
f(r_{\rm H})=0.
\end{eqnarray}
while the horizon angular velocity is given by
\begin{eqnarray}
\Omega_{\rm H} = \frac{\omega(r_{\rm H})}{r_{\rm H}}.
\label{Omega}
\end{eqnarray}

We focus our study on extremal black holes. 
In quasi-isotropic coordinates, the event horizon of extremal black holes is given by
$r_{\rm H}=0$. Note that in this case, the horizon angular velocity is obtained from
\begin{eqnarray}
\Omega_{\rm H} = \lim_{r_{\rm H} \rightarrow 0}\frac{\omega(r_{\rm H})}{r_{\rm H}} = \omega'(r)|_{\cal H}.
\end{eqnarray}

The area of the horizon $A_{\rm H}$ is given by
\begin{equation}
A_{\rm H}=\int_{{\cal H}} \sqrt{|g^{(3)}|}=r_{\rm H}^{3} A(S^{3}) \lim_{r \to r_{\rm H}}
 \sqrt{\frac{m^{2} n}{f^{3}}} \label{hor_area} , \end{equation}
and the surface gravity $\kappa$ reads
\begin{equation}
\kappa^2 = -\frac{1}{2}(\nabla\zeta)^2|_{\cal H}
 =  \lim_{r \to r_{\rm H}} \frac{f}{(r-r_{\rm H}) \sqrt{m}}
 . \label{kappa} \end{equation}
For extremal black holes the surface gravity vanishes, $\kappa=0$.

The horizon mass $M_{\rm H}$ and horizon angular momenta
$J_{{\rm H} (k)}$ are given by
\begin{equation}
M_{\rm H} = - \frac{3}{2} \int_{{\cal H}} \alpha 
 , \label{Hmass} \end{equation}
\begin{equation}
J_{{\rm H} (k)} =   \int_{{\cal H}} \beta_{(k)} \ 
 . \label{Hang} \end{equation}
For equal-magnitude angular momenta
$J_{{\rm H} (k)} = \varepsilon_k J_{\rm H}$, $k=1, 2$.

The electric charge $Q$ can also be determined at the horizon,
\begin{equation}
Q= - \frac{1}{2} \int_{{\cal H}}  
\left( \tilde F 
+\frac{\lambda}{\sqrt{3}} A \wedge F \right)
 . \label{Hcharge} 
\end{equation}
The horizon electrostatic potential $\Phi_{\rm H}$ is defined by
\begin{equation}
\Phi_{\rm H} = \left. \zeta^\mu A_\mu \right|_{\cal H}
\ =\left. (a_{0}+\Omega_{\rm H} a_{\varphi} )\right|_{\cal H} . \label{Phi} 
\end{equation}
Note that $\Phi_{\rm H}$ is constant at the horizon.

\subsection{Scaling Symmetry}
The EMCS solutions have the following scaling
symmetry \cite{Kunz:2005ei}
( with $\tau>0$ an arbitrary parameter)
\begin{equation}
 \tilde M= \tau^{2} M , \ \ \ 
 \tilde J_i= \tau^{3} J_i , \ \ \ 
 \tilde r_{\rm H}=\tau r_{\rm H}  ,\ \ \ 
 \tilde \Omega_{\rm H} = \Omega_{\rm H}/\tau  ,\ \ \ 
 \tilde \kappa = \kappa/\tau  ,
\label{scaling} 
\end{equation}
etc.
Let us therefore introduce scaled quantities,
where we scale with respect to appropriate powers
of the mass. These scaled quantities include
the scaled angular momentum
$j=J/M^{3/2}$,
the scaled charge $q=Q/M$,
the scaled area $a_{\rm H} = A_{\rm H}/M^{3/2}$,
the scaled surface gravity $\bar \kappa = \kappa M^{1/2}$,
and the scaled horizon angular velocity $\bar \Omega_{\rm H} = \Omega_{\rm H} M^{1/2}$.

\subsection{Smarr Formula}

The Smarr mass formula for EMCS black holes
with two equal-magnitude angular momenta 
reads  \cite{Gauntlett:1998fz}
\begin{equation}
 M = 3 \kappa A_{\rm H} + 3 \Omega_{\rm H}
J  + \Phi_{\rm H} Q  \ , \label{smarr}
\end{equation}
or in terms of scaled quantities
\begin{equation}
1= 3\bar \kappa a_{\rm H} + 3 \bar \Omega_{\rm H} j +  \Phi_{\rm H}q \ . \label{mass3}
\end{equation}
Note that in the extremal case, the surface gravity vanishes.    
Moreover, in the extremal case
\begin{equation}
M_{\rm H} = 3\Omega_{\rm H} J_{\rm H}.
\end{equation} 

EMCS black holes satisfy the first law \cite{Gauntlett:1998fz}
\begin{equation}
\label{first-law}
dM = 2 \kappa\, dA_{\rm H} + 2 \Omega_{\rm H}\, dJ+\Phi_{\rm H}\, dQ. 
\end{equation}

\section{ Extremal configurations: near-Horizon Solutions}  

Before analyzing the numerical solutions, 
some analytical understanding of the properties 
of extremal EMCS black holes can be achieved by
deriving the near-horizon solutions
in the entropy function formalism
\cite{Sen:2005wa,Astefanesei:2006dd,Goldstein:2007km}.
Employing this formalism we obtain semi-analytic 
expressions for the entropy 
as a function of the electric charge and the angular momentum. 
We note, however, that the near horizon formalism needs special care, 
when it is employed in the presence of a CS term, 
as discussed extensively in the literature 
(see e.g.~\cite{Suryanarayana:2007rk}).

In the following we first review the general entropy function formalism
in the presence of a CS term. Subsequently, we apply the formalism
to construct the near-horizon solutions 
of the extremal black holes in (i) pure Einstein-Maxwell theory,
(ii) the bosonic sector of $D=5$ supergravity,
and (iii) EMCS theory with general CS coupling constant $\lambda$.

\subsection{Entropy function formalism}

To apply the entropy function formalism to obtain the near-horizon
geometry of extremal EMCS solutions, we make use of the Ansatz 
\begin{eqnarray}
\label{metric_ansatz}
ds^2 &=& v_1(\frac{dr^2}{r^2}-r^2dt^2) + v_2[4d\theta^2+\sin^22\theta(
\varepsilon_2 d\varphi_2- \varepsilon_1 d\varphi_1)^2]
\\ 
\nonumber
&+&v_2\eta[\varepsilon_1 d\varphi_1+
\varepsilon_2 d\varphi_2+\cos{2\theta}
(\varepsilon_2 d\varphi_2- \varepsilon_1 d\varphi_1)-\alpha r dt]^2.
\end{eqnarray}
Note the shift of the radial coordinate $r\rightarrow r - r_{\rm H}$. Thus
the horizon is located at $r=0$.

For the gauge potential we employ the Ansatz
\begin{eqnarray}
\label{gauge_ansatz}
A &=& -(\rho + p\,\alpha)rdt + 2p\,(\sin^2\theta \varepsilon_1 d\varphi_1
+ \cos^2\theta \varepsilon_2 d\varphi_2).
\end{eqnarray}
Here the parameters $v_1$, $v_2$, $\eta$, $\alpha$, $\rho$ and $p$ 
are constants, which satisfy a set of algebraic relations, 
that follow within the near-horizon formalism 
\cite{Sen:2005wa,Astefanesei:2006dd,Goldstein:2007km}.
 
In the near-horizon formalism, the entropy is obtained from the extremum
of the entropy function 
\begin{equation}
S=2\pi (2 \alpha J + \rho\, \hat q-h) , 
\label{entropy_function}
\end{equation}
where $h$ is an action functional,
that depends on the constants of the near horizon Ansatz
\begin{eqnarray}
\label{at2}
\nonumber
h(\alpha,v_1,v_2,\eta,p,\rho)
&=&\int d \theta d \varphi_1 d\varphi_2 \sqrt{-g} {\cal L}\\\nonumber
&=&\frac{8\pi^2}{9v_1\sqrt{v_2}}(
9v_2^3\eta^{3/2}\alpha^2-9v_2v_1^2\eta^{3/2}-36v_2^2\sqrt\eta v_1 + 36v_1^2\sqrt\eta v_2 + 36v_2^2\sqrt\eta \rho^2 \\
&+& 72v_2^2\sqrt \eta \rho p \alpha + 36v_2^2\sqrt\eta p^2 \alpha^2 - 36 v_1^2\sqrt\eta p^2 - 32\lambda\sqrt 3 p^2 \rho v_1 \sqrt v_2 - 32 \lambda \sqrt 3 p^3 \alpha v_1 \sqrt v_2) ,
\end{eqnarray}
and $\hat q$ is related to the electric charge (see Eq.~(\ref{q_hat})).

In the usual analysis, one proceeds by taking the derivative
of the functional $h$ with respect to the constants
introduced in the Ansatz.
These then yield the set of equations to be solved and the conserved charges
$J$ and $Q$.
However, in the presence of a Chern-Simons term in the action
the analysis needs to be modified \cite{Suryanarayana:2007rk}.

To obtain the near-horizon geometry we can directly 
solve the Einstein equations and the Maxwell equations.
The set of Einstein equations is equivalent to taking 
\begin{eqnarray}
\label{at3}
 \frac{\partial h}{\partial v_1}=0, \ \ \
 \frac{\partial h}{\partial v_2}=0, \ \ \
 \frac{\partial h}{\partial \eta}=0,
\end{eqnarray}
$i.e.$, these equations continue to hold also for EMCS theory.
Variations of the functional with respect to $p$,
however, do not take into account the Chern-Simons contribution. 
Therefore the Maxwell equations must be used when
$\lambda\neq 0$. Simply setting $\partial h/\partial p = 0$
 would lead to a wrong result in this case.

The Einstein equations then yield the following algebraic relations
\begin{eqnarray}
\label{relgen}
v_2 &=& v_1, \nonumber\\
\eta v_1 &=& -\frac{4}{3}\frac{(\rho-p+p\alpha)(\rho+p+p\alpha)}{\alpha^2-1}, \nonumber\\
v_1 &=& \frac{2}{3}\frac{\alpha^4p^2-p^2+2\alpha^3\rho p-4\rho p \alpha + \alpha^2\rho^2 - 2\rho^2}{\alpha^2-1},
\end{eqnarray}
and the Maxwell equations lead to
\begin{eqnarray}
\label{relmax}
3\alpha v_2^{5/2}\sqrt\eta\rho + 3\alpha^2v_2^{5/2}\sqrt\eta p - 4\lambda\sqrt 3 p v_1 v_2 \rho - 4 \lambda \sqrt 3 p^2 v_1 v_2 \alpha - 3p v_1^2\sqrt v_2 \sqrt \eta = 0.
\end{eqnarray}
Note, that we  now have 4 algebraic relations for 6 unknown constants. 
Thus we have two independent parameters. 
For those we may choose the angular momentum $J$ 
and the electric charge $Q$.

In the usual near-horizon formalism, the angular momentum $J$ is obtained
by taking the derivative of the action functional $h$ with respect
to the associated constant $\alpha$,
and the electric charge $Q$ is obtained by
taking the derivative with respect to $\rho$.
To see where modifications arise because of the CS term
we now recall the expressions for the angular momentum
and the electric charge as Noether charges
\cite{Wald:1993nt,Lee:1990nz,Rogatko:2007pv,Suryanarayana:2007rk}.

The Noether charges can be obtained by integrating the corresponding
charge densities \cite{Suryanarayana:2007rk}
\begin{eqnarray}
\label{Noether_J}
Q_{\xi}^{\alpha \mu} = -\left[\sqrt{-g}\left(\nabla^{\alpha}\xi^{\mu} - \nabla^{\mu}\xi^{\alpha}\right) - 4(\xi^{\tau}A_{\tau})(\sqrt{-g}F^{\alpha\mu} + \frac{2\lambda}{3\sqrt 3}\epsilon^{\alpha \mu \beta \nu \rho}A_{\beta}F_{\nu \rho})\right]
\end{eqnarray}
over the $S^3$-sphere.
When taking the Killing vector $\eta_{(1)}$ or $\eta_{(2)}$, 
the angular momentum $J$ is obtained 
\begin{eqnarray}
J = \int d \theta d \varphi_1 d\varphi_2 Q_{\xi_{\varphi_1}}^{t r} = 64\pi^2\frac{v_2^{3/2}}{v_1}\sqrt \eta p(\rho+p\alpha) + 16\pi^2\frac{v_2^{5/2}}{v_1}\eta^{3/2}\alpha - \frac{256}{9}\sqrt 3 \pi^2p^3\lambda.
\end{eqnarray}
Since the very same result is obtained from the equation
\begin{eqnarray}
 \frac{\partial h}{\partial \alpha}&=&J ,
\end{eqnarray}
the usual near-horizon relation remains valid
for the angular momentum.

However, this is not true for the electric charge $Q$. 
For $Q$ we need to consider the charge density \cite{Suryanarayana:2007rk}
\begin{eqnarray}
\label{Noether_Q}
Q_M^{\alpha \mu} = 4\left[\sqrt{-g}F^{\alpha\mu} + \frac{\lambda}{\sqrt 3}\epsilon^{\alpha \mu \beta \nu \rho}A_{\beta}F_{\nu \rho}\right].
\end{eqnarray}
When we integrate this charge density over the $S^3$-sphere we obtain
the charge $Q$
\begin{eqnarray}
Q = \int d \theta d \varphi_1 d\varphi_2 Q_M^{t r} = -64\pi^2\frac{v_2^{3/2}}{v_1}\sqrt \eta (\rho+p\alpha) + \frac{128\pi^2\sqrt 3}{3}\lambda p^2.
\end{eqnarray}
Note, that this is equivalent to
\begin{eqnarray}
 \frac{\partial h}{\partial \rho}&=&\hat q= -Q - \frac{128\pi^2\sqrt
   3}{9}\lambda p^2, \label{q_hat}
\end{eqnarray}
and thus ${\partial h}/{\partial \rho} \ne -Q$,
in contrast to a naive application of the entropy function formalism.
Consequently, 
the extremization of the entropy functional must be appropriately
modified to obtain the proper set of equations and charges.

We are also interested in obtaining expressions for the horizon properties.
The horizon angular momentum can be calculated from the standard Komar formula
\begin{eqnarray}
J_{\rm H} = \int d \theta d \varphi_1 d\varphi_2 (-\sqrt{-g})(\nabla^{t}\xi^{r} - \nabla^{r}\xi^{t}) 
=  16\pi^2\frac{v_2^{5/2}}{v_1}\eta^{3/2}\alpha.
\end{eqnarray}
Finally, for the horizon area we find the following expression
\begin{eqnarray}
A_{\rm H} = \int d \theta d \varphi_1 d\varphi_2 \sqrt{|\det(g^{(3)})|} 
= 16\pi^2v_2^{3/2}\sqrt \eta.
\end{eqnarray}

 Unfortunately, for a generic nonzero CS coupling constant $\lambda$,
$\lambda \neq \lambda_{\rm SG}$, 
it is not possible to give an explicit expression for $S=A_{\rm H}/4G_5$  as a
function of $Q$ and $J$. 
A straightforward numerical analysis of the algebraic relations
reveals a rather complicated picture,
with several branches of solutions.
We are now going to describe several cases in detail.

\boldmath
\subsection{Einstein-Maxwell: $\lambda_{\rm EM}=0$}
\unboldmath

In the pure EM case the CS term vanishes, since $\lambda=0$,
and the above
relations reduce to those obtained in the usual entropy formalism.
The EM case is interesting, since it allows for 
two branches of near-horizon solutions
\cite{Blazquez-Salcedo:2013yba,Kunduri:2013gce,Blazquez-Salcedo:2013wka}.

The near-horizon branch starting from the Myers-Perry solution is given by
\begin{eqnarray}
&&v_1 = v_2  ,\nonumber \\
&&\eta = \frac{2(v_2-2p^2)}{v_2}  ,\nonumber \\
&&\rho = 0  ,\nonumber \\
&&\alpha = -1 , \\
&&J = 32\pi^2 v_2^{3/2} \sqrt{2-4p^2/v_2}  , \nonumber \\
&&Q = 64\pi^2 p \sqrt v_2 \sqrt{2-4p^2/v_2}  ,\nonumber 
\end{eqnarray}
where $J$ and $Q$ are the 
angular momentum and the electric charge, respectively.
The horizon angular momentum $J_{\rm H}$ and horizon area $A_{\rm H}$ are given by
\begin{eqnarray}
&&J_{\rm H} = 16\pi^2v_2^{3/2}(2-4p^2/v_2)^{3/2}  ,\nonumber \\
&&A_{\rm H} = 16\pi^2 v_2^{3/2} \sqrt{2-4p^2/v_2} = J/2  . 
\end{eqnarray}
Hence, along the full Myers-Perry branch,
the entropy is always proportional to the angular momentum
and independent of the charge.

 A second  near-horizon branch 
starts from the Reissner-Nordstr\"om solution,
and is  given by

\begin{eqnarray}
&&v_1 = v_2=\frac{4(\alpha^2+1)\rho^2}{3(\alpha^2-1)^2}  ,\nonumber \\
&&\eta = \frac{1}{\alpha^2+1}  ,  \nonumber \\
&&p = -\frac{\rho\alpha}{\alpha^2-1} , \\
&&J = \frac{512\sqrt 3 \pi^2 \rho^3\alpha}{9(\alpha^2-1)^3}  , \nonumber \\
&&Q = \frac{128\sqrt 3 \pi^2 \rho^2}{3(\alpha^2-1)^2}  . \nonumber
\end{eqnarray}
It has horizon angular momentum and horizon area 
\begin{eqnarray}
&&J_{\rm H} = \frac{128\sqrt 3 \pi^2 \rho^3\alpha}{9(\alpha^2-1)^3} = \frac{J}{4}  ,
\nonumber \\
&&A_{\rm H} = \frac{128\sqrt 3 \pi^2 \rho^3(\alpha^2+1)}{9(\alpha^2-1)^3} 
= \frac{\sqrt 2 \pi 3^{3/4}}{2}J^2Q^{-3/2} + \frac{3^{1/4}\sqrt
  2}{48\pi}Q^{3/2}  .
\end{eqnarray}
Along this branch, the area is not proportional to the angular momentum. 
Instead, the horizon angular
momentum is proportional to the total angular momentum.
Interestingly, only parts of the near-horizon branches are
also realized as global solutions
\cite{Blazquez-Salcedo:2013yba,Blazquez-Salcedo:2013wka}.

\boldmath
\subsection{$D=5$ supergravity: $\lambda_{\rm SG}=1$}
\unboldmath

In the case of $D=5$ supergravity, $\lambda_{\rm SG}=1$, 
the global solutions are known \cite{Chong:2005hr},
with the BMVP solution representing the special case
of supersymmetric and thus ergofree solutions
\cite{Breckenridge:1996is}.
It is straightforward to obtain the near-horizon solutions
in this case. 
They can be expressed as \cite{Suryanarayana:2007rk}
\begin{eqnarray}
&&v_1 = v_2 = \mu/4  , \nonumber \\
&&\eta = 1 - \frac{j^2}{\mu^3}  , \nonumber \\
&&\alpha = \frac{j}{\sqrt{\mu^3-j^2}}  , \nonumber \\
&&\rho = -\frac{\sqrt 3 \mu^2}{4\sqrt{\mu^3-j^2}}  ,  \\
&&p = \frac{\sqrt 3}{4}\frac{j}{\mu}  , \nonumber \\
&&J = -4\pi^2 j  , \nonumber \\
&&Q = 8\sqrt{3} \pi^2\mu ,\nonumber
\end{eqnarray}
where the parameters $\mu$ and $j$ essentially describe
the electric charge and the total angular momentum.

The horizon angular momentum and the horizon area are given by
\begin{eqnarray}
&&J_{\rm H} = \frac{2\pi^2 j}{\mu^3}(\mu^3-j^2)  , \\
&&A_{\rm H} = 2\pi^2\sqrt{\left|\mu^3-j^2\right|} 
= \frac{\sqrt 2}{48\pi}\sqrt{\left|\sqrt 3 Q^3 - 288\pi^2 J^2 \right|} \nonumber  .
\end{eqnarray}
This shows, that there are again two branches of near-horizon solutions.
The first branch has 
$J^2 > - \frac{4}{3 \sqrt{3} \pi} Q^3$,
while the second branch has
$J^2 < - \frac{4}{3 \sqrt{3} \pi} Q^3$.
The second branch thus corresponds to the BMPV branch.
In terms of the global solutions, this branch is ergo-region free
with vanishing horizon angular velocity.
In contrast, the first branch
corresponds to the ordinary branch which possesses an ergo-region.
At the matching point of both branches the area vanishes.

\boldmath
\subsection{Generic values of $\lambda$}
\unboldmath

Let us now consider the near-horizon solutions for generic 
values of the CS coupling constant $\lambda$. 
Note, that the solutions cannot be given explicitly,
except for the previous two cases.
Also, for $\lambda \ne 0$
the CS term breaks the charge reversal symmetry $Q \rightarrow -Q$ 
for spinning solutions.
Therefore, we need to consider positive and negative $Q$
separately for finite values of $\lambda$.

Let us first consider near-horizon solutions with positive $Q$.
When $\lambda$ is increased from zero, 
the branch structure of the pure EM case is lost. 
Instead a new branch structure emerges,
which is generic for $0 < \lambda < 1.91$.
It resembles the supergravity case,
which is included as a particular case.
After a small transition region $1.91 < \lambda < 2$
a new generic branch structure arises for $\lambda > 2$.

\begin{figure}[h!]
\mbox{\hspace*{-1.0cm}
\subfigure[][]{
\includegraphics[height=.43\textheight, angle
=270]{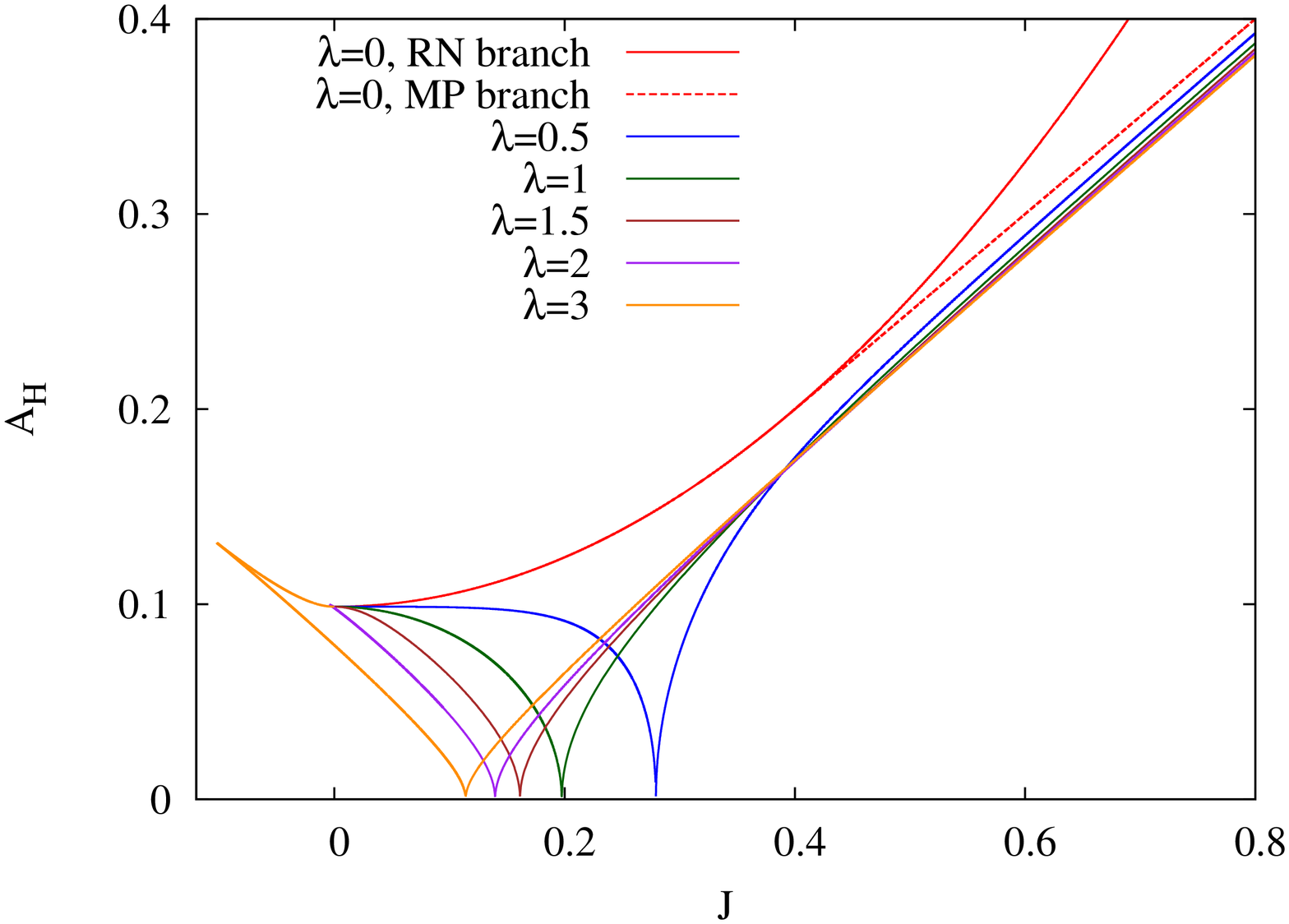}
\label{plot_J_Ah_attractor_different_lambda} 
}
\hspace*{-0.6cm}
\subfigure[][]{
\includegraphics[height=.39\textheight,
angle=270]{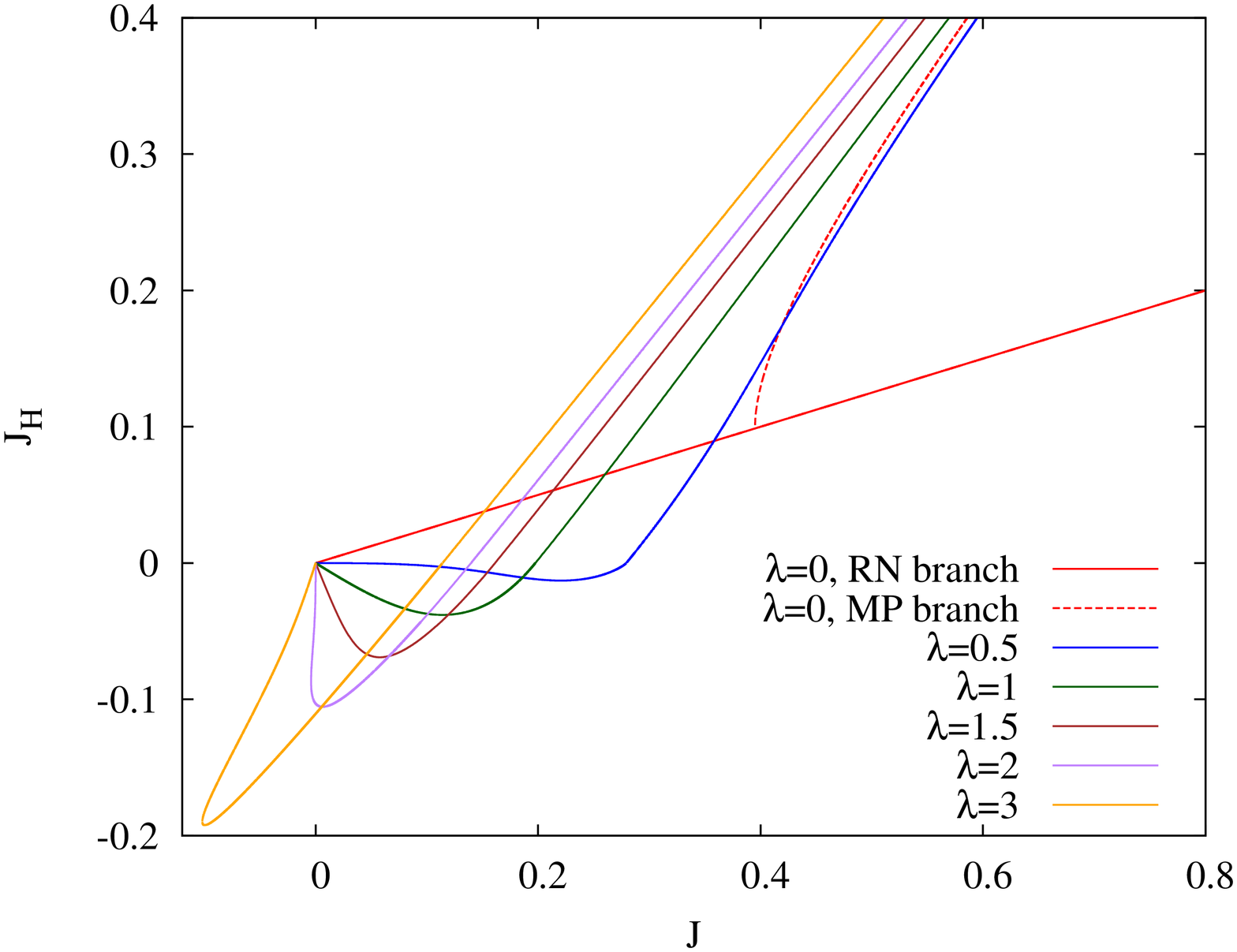} 
\label{plot_J_Jh_attractor_different_lambda} 
}
}
\caption{\small
Near-horizon solutions:
The horizon area $A_{\rm H}$ (a) and the horizon angular momentum $J_{\rm H}$ (b)
versus the angular momentum $J$ 
for CS coupling constant $\lambda=0, \ 0.5, \ 1, \ 1.5, \ 2,$ and $3$ 
and charge $Q = 4$.
The symmetric solutions obtained for $J \rightarrow -J$ are suppressed.
}
\label{fig1}
\end{figure}

Let us now demonstrate this $\lambda$-dependence in detail.
In Fig.~\ref{plot_J_Ah_attractor_different_lambda} 
we exhibit the horizon area $A_{\rm H}$ versus
the angular momentum $J$ for several values of $\lambda$,
where $\lambda$ is increased from zero to three 
and the charge is fixed at the positive value $Q=4$.
Note, that the set of symmetric solutions obtained for $J \rightarrow -J$ 
is not exhibited here.
It is neither exhibited in
Fig.~\ref{plot_J_Jh_attractor_different_lambda}, which
shows the horizon angular momentum $J_{\rm H}$ versus
$J$ for the same set of parameters.

Keeping in mind that a set of symmetric solutions 
is obtained for $J \rightarrow -J$,
we focus for the moment the discussion on the solutions exhibited in 
the figure.
For $\lambda \neq 0$ and positive $Q$, there exists always
a solution with vanishing area.
The angular momentum $J$ of this solution is
finite. For fixed $Q$ its angular momentum
decreases monotonically with increasing $\lambda$,
as seen in Fig.~\ref{plot_J_Ah_attractor_different_lambda}.
Fig.~\ref{plot_J_Jh_attractor_different_lambda}
demonstrates, that the horizon angular momentum changes smoothly
at these solutions with vanishing area.

In the interval $0 < \lambda < 1.91$ 
there are two near-horizon branches.
The small-$J$ branch
extends from the extremal Reissner-Nordstr\"om solution
at $J=0$ to the solution with vanishing area,
which has the maximal value of $J$ along this branch.
The large-$J$ branch, on the other hand, extends from 
the solution with vanishing area and the minimal value of $J$
along this branch to solutions with arbitrarily large values of $J$
(Myers-Perry limit).
Thus the structure of the near-horizon branches is
essentially the same as the one found 
for the supersymmetric value $\lambda_{\rm SG}=1$,
except that for $\lambda_{\rm SG}=1$ the small-$J$ branch is ergo-region
free.

\begin{figure}[h!]
\begin{center}
\mbox{\hspace*{-1.0cm}
\subfigure[][]{
\includegraphics[height=.4\textheight, angle
=270]{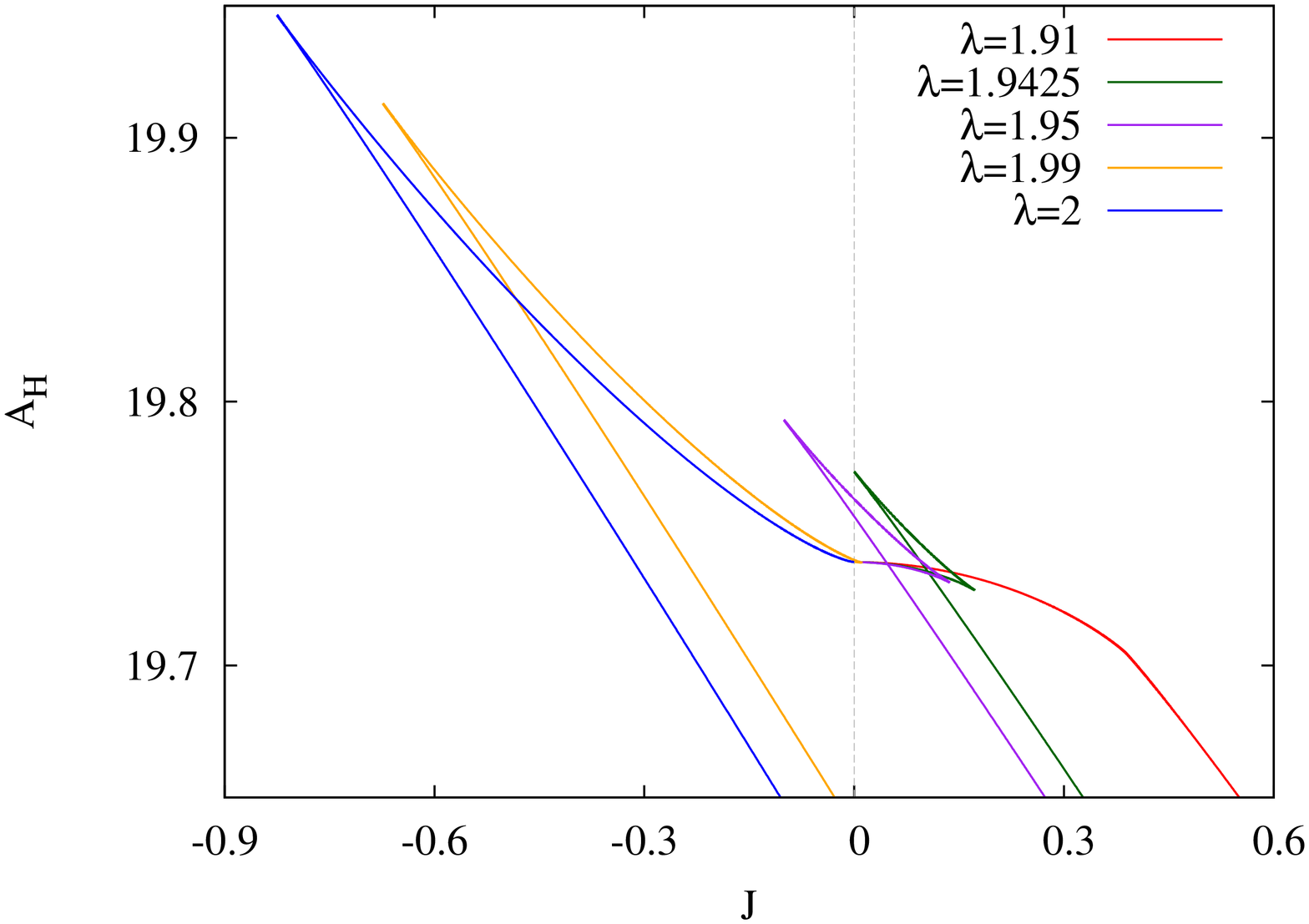} 
\label{plot_J_Ah_attractor_lambda_near_2_newQ} 
}
\hspace*{-0.6cm}
\subfigure[][]{
\includegraphics[height=.4\textheight,
angle=270]{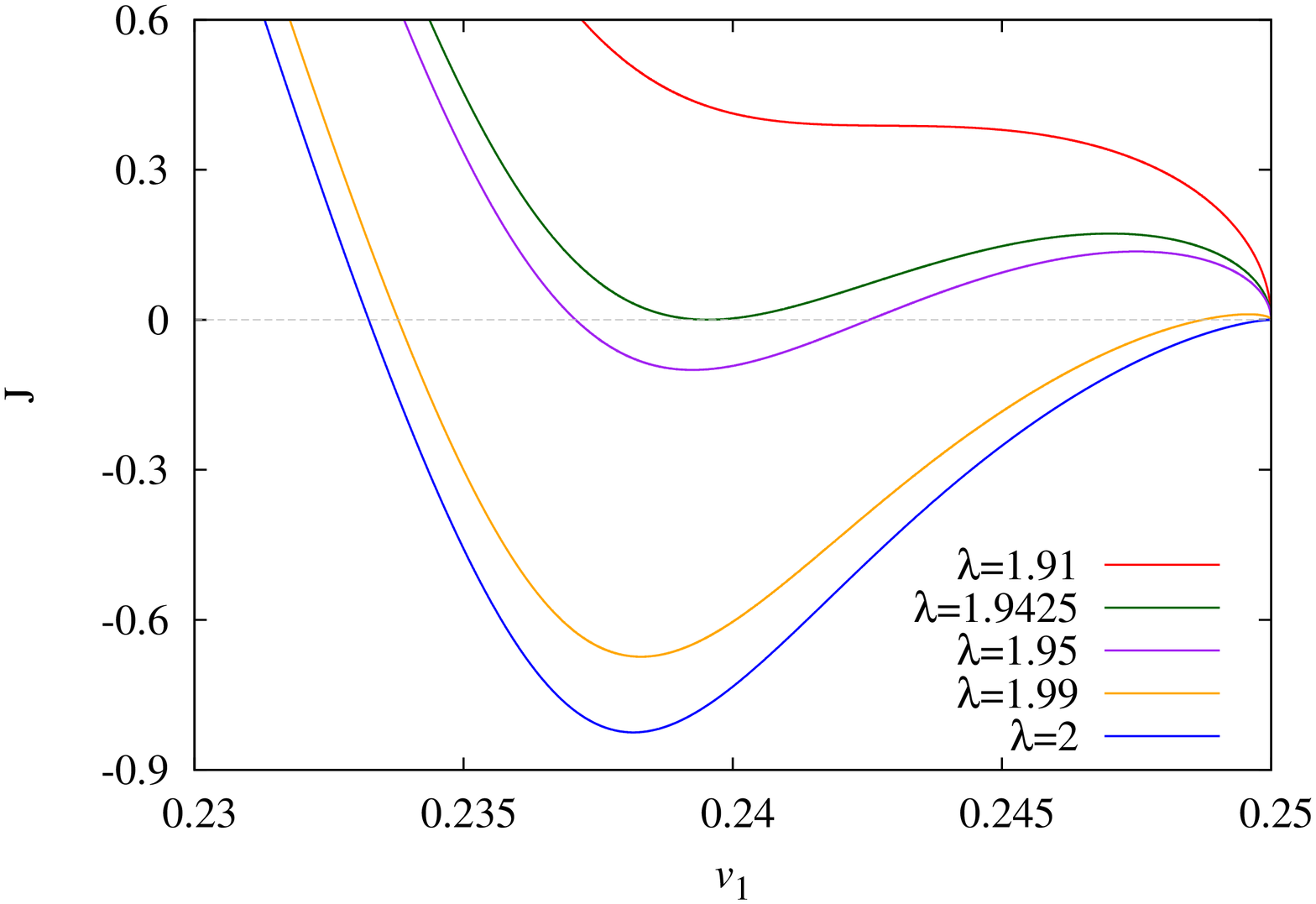} 
\label{plot_v1_J_attractor_lambda_near_2_newQ} 
}
}
\end{center}
\vspace*{-0.5cm}
\caption{\small
Near-horizon solutions:
(a) The area $A_{\rm H}$ versus the angular momentum $J$ 
for values of the CS coupling covering the vicinity of the
transition region near the critical value $\lambda_{\rm cr}=2$. 
The value of the charge is $Q=8\sqrt{3}\pi^2$. 
The symmetric solutions for $J \rightarrow -J$ are not shown.
(b) The angular momentum $J$ 
versus the parameter $v_1$ for the same sets of solutions.}
\label{fig2}
\end{figure}

Let us next address the transition region $1.91 < \lambda < 2$.
Here the branch structure changes. This is illustrated
in Figs.~\ref{plot_J_Ah_attractor_lambda_near_2_newQ} and
\ref{plot_v1_J_attractor_lambda_near_2_newQ}. 
Note, that the set of symmetric solutions obtained for $J \rightarrow -J$ 
is again not exhibited here to have more clarity in the figures.

When the horizon area $A_{\rm H}$
is considered as a function of the angular momentum $J$,
as shown in Fig.~\ref{plot_J_Ah_attractor_lambda_near_2_newQ},
a swallowtail bifurcation is seen
for any value of $\lambda$ in
the transition region $1.91 < \lambda < 2$.
Hence, there are solutions with the same
value of $J$ but different values of the horizon 
area or the horizon angular momentum
in the transition region.
The reason for the emergence of the swallowtail bifurcation
becomes clear by inspecting
Fig.~\ref{plot_v1_J_attractor_lambda_near_2_newQ},
where the angular momentum is shown versus
the near-horizon parameter $v_1$.
Here an inflection point arises at $\lambda = 1.91$,
and entails the occurrence of a local minimum and a local maximum
for larger values of $\lambda$.

With increasing $\lambda$ the 
endpoints of the two cusps of the swallowtail bifurcation
move to smaller
values of $J$ and separate from each other.
This has two interesting effects.
The first effect is that, at the particular value of $\lambda=1.9425$,
the upper cusp endpoint reaches the value $J=0$. 
The associated black hole solution thus corresponds
to an extremal stationary black hole,
that is not static, $i.e.$, it differs from
the extremal static Reissner-Nordstr\"om solution,
that is also present.
(Of course, by the symmetry $J \rightarrow - J$
there are two such stationary nonstatic solutions.
These have the same area but
differ in the signs of their horizon angular velocities
and horizon angular momenta.)

As seen in Fig.~\ref{plot_J_Ah_attractor_lambda_near_2_newQ},
for $1.9425 < \lambda < 2$ 
the upper cusp endpoint is at negative values of $J$
and thus the number of nonstatic $J=0$ solutions is doubled.
With increasing $\lambda$, the upper cusp endpoint continues
to move to lower values of $J$, and so does the lower cusp endpoint.
However, for any value of $\lambda$, the extremal Reissner-Nordstr\"om
solution also resides at $J=0$,
and is thus approached by the lower cusp endpoint.
This leads to the second interesting effect,
namely the disappearance of the lower cusp
at the critical value $\lambda_{\rm cr}=2$,
and thus the disappearance 
of one of these (two pairs of ) $J=0$ solutions.

In particular, as seen in 
Fig.~\ref{plot_v1_J_attractor_lambda_near_2_newQ},
when the critical value $\lambda_{\rm cr}=2$ is
approached, the local minimum in $J$ deepens further, 
while the local maximum merges with the endpoint 
corresponding to the static solution.
From the point of view of the full set of solutions,
$i.e.$, including the symmetric $J \rightarrow -J$ set of solutions,
at the critical value $\lambda_{\rm cr}=2$
two swallowtail bifurcations merge to form
a single swallowtail bifurcation.
Thus at $\lambda_{\rm cr}=2$ in total only one pair of nonstatic
$J=0$ solutions remain together with the static 
Reissner-Nordstr\"om solution.

\begin{figure}[h!]
\begin{center}
\mbox{\hspace*{-1.0cm}
\subfigure[][]{
\includegraphics[height=.4\textheight, angle
=270]{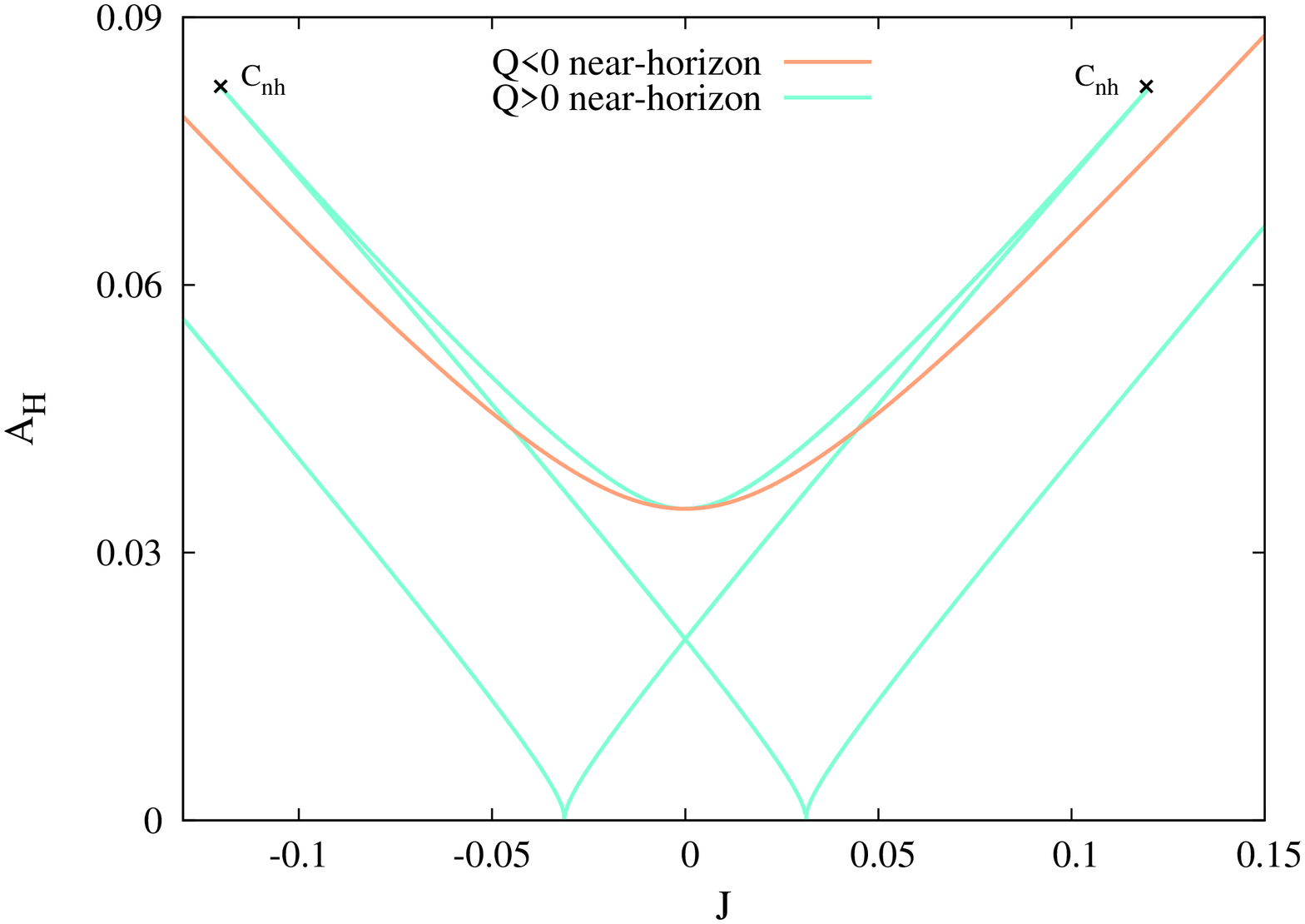}
\label{plot_J_Ah_lambda_5_extremal_long_paper_NH_ver.ps}
}
\hspace*{-0.6cm}
\subfigure[][]{
\includegraphics[height=.4\textheight,
angle=270]{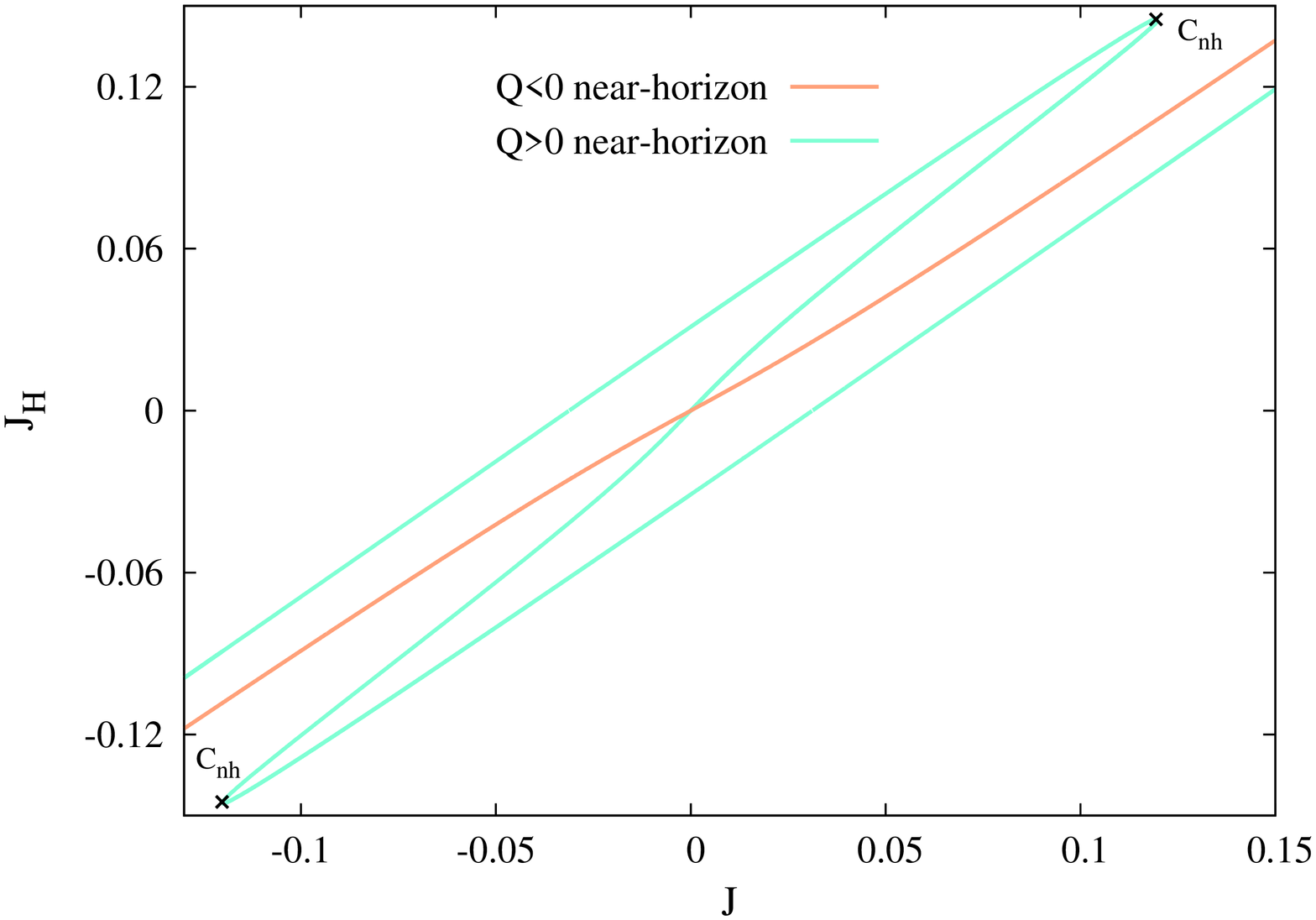} 
\label{plot_J_Jh_lambda_5_extremal_long_paper_NH_ver.ps}
}
}
\end{center}
\vspace*{-0.5cm}
\caption{\small
Near-horizon solutions:
The horizon area $A_{\rm H}$ (a) and
horizon angular momentum $J_{\rm H}$ (b)
versus the angular momentum $J$ 
for CS coupling constant $\lambda=5$. 
The value of the charge is $|Q|=1$.
}
\label{fig3}
\end{figure}

Beyond the critical value $\lambda_{\rm cr}=2$ 
we observe a new generic branch structure,
exhibited in Figs.~\ref{plot_J_Ah_lambda_5_extremal_long_paper_NH_ver.ps}
and \ref{plot_J_Jh_lambda_5_extremal_long_paper_NH_ver.ps}.
These figures show the full set of near-horizon solutions,
including the symmetric set of solutions 
obtained for $J \rightarrow -J$
as well as the solutions for negative values of the charge.
In the figures the value $\lambda=5$ is chosen
for the CS coupling constant.

For positive $Q$
these near-horizon solutions possess three distinct branches,
while they feature only a single branch for negative $Q$.
Because of the symmetry $J \rightarrow -J$,
all these branches have mirror branches in
Fig.~\ref{plot_J_Ah_lambda_5_extremal_long_paper_NH_ver.ps}. 
Considering first positive $Q$,
the small-$J$ branch starts from the extremal
Reissner-Nordstr\"om solution and extends until
the cusp $C_{\rm nh}$ is reached. 
The intermediate-$J$ branch extends from the cusp $C_{\rm nh}$
to the solution with vanishing area.
From there the large-$J$ branch extends towards solutions with
arbitrarily large values of the angular momentum,
reaching in the limit $J \to \infty$ the Myers-Perry solution.

Clearly, beside the static Reissner-Nordst\"om solution
there are two symmetric $J=0$ solution, which are nonstatic.
As seen in Fig.~\ref{plot_J_Jh_lambda_5_extremal_long_paper_NH_ver.ps}
their horizon angular momenta are finite and opposite to each other.
These nonstatic $J=0$ solutions are only present in the set of
positive $Q$ solutions. 
Likewise the solutions with vanishing area 
are only present in the set of
positive $Q$ solutions.
In contrast, the negative $Q$ solutions 
show a rather unspectacular behavior.
Their single branch connects directly from the
Reissner-Nordstr\"om solution to the Myers-Perry solution 
for any finite value of the CS coupling constant $\lambda$.

Concluding, we have seen
that the branch structure of the near-horizon solutions
depends significantly on the value of $\lambda$
and on the sign of the charge $Q$.
In the next section we discuss
the numerically obtained sets of global solutions
and compare with the corresponding sets of near-horizon solutions.
We show that for negative $Q$,
all near-horizon solutions correspond to global solutions.
In contrast, when positive values of $Q$ are considered,
not all of the near-horizon solutions are realized globally.
Moreover, other near-horizon solutions may correspond to
more than a single global solution.


\section{Global solutions: Numerical procedure}

Having obtained analytical expressions for the extremal solutions
in the near-horizon formalism, we are now interested in obtaining global solutions,
both extremal and non-extremal.
 As discussed in Section II, 
for generic values of the Chern-Simons coupling constant $\lambda$ and
of the global charges, numerical integration of the differential
equations seems necessary.

In the following we discuss the numerical procedure, the boundary
conditions and the expansions of the functions
employed to obtain generic EMCS black holes in five dimensions.

\subsection{Differential equations and boundary conditions}

Considering variation of the Lagrangian with respect to the $a_0(r)$ component 
of the gauge field we can obtain a first integral of the system in
terms of the electric charge of the black hole.
%
\begin{eqnarray}
\frac{d}{dr}a_0(r) = - \frac{\omega(r)}{r}\frac{d}{dr}a_{\varphi}(r)
+ \frac{f(r)^{3/2}}{\sqrt{m(r)n(r)}}\left[ \frac{4}{3}\sqrt{3}\lambda a_{\varphi}(r)^2 
- \frac{Q}{2} \right].
\end{eqnarray}
This differential equation is compatible with the
Einstein and Maxwell equations. Combining this first-order differential
equation with the system of differential equations, we obtain a minimal system
of differential equations for the EMCS black hole solutions: 
four second-order
differential equations for $f(r)$, $m(r)$, $\omega(r)$, and $a_{\varphi}(r)$, one
first-order differential equation for $n(r)$, and one first-order differential
equation for $a_0(r)$, which is decoupled from the other differential
equations.

 For the numerical calculations, we have found it useful to introduce
a compactified radial coordinate. For the non-extremal solutions we take the
compactified coordinate to be
$x= 1-r_{\rm H}/r$. In the extremal case we employ $x= \frac{r}{1+r}$.
(Note, that we are using an isotropic radial coordinate $r$, so $r_{\rm H}=0$ in
the extremal case.)
We employ a collocation method for boundary-value ordinary
differential equations, equipped with an adaptive mesh selection procedure
\cite{COLSYS}.
Typical mesh sizes include $10^3-10^4$ points.
The solutions have a relative accuracy of $10^{-10}$.
The estimates of the relative errors of the global charges
and the magnetic moment are of order $10^{-6}$, 
giving rise to an estimate of the relative error of 
the gyromagnetic ratio $g$ of order $10^{-5}$. 

To obtain asymptotically flat solutions
the metric functions should  satisfy the boundary conditions at infinity
\begin{equation}
f|_{r=\infty}=m|_{r=\infty}=n|_{r=\infty}=1 \ , \ \omega|_{r=\infty}=0 
\ . \label{bc1} \end{equation}
For the gauge potential we choose a gauge  in which it vanishes
at infinity
\begin{equation}
a_0|_{r=\infty}=a_\varphi|_{r=\infty}=0 
\ . \label{bc2} \end{equation}

Requiring the horizon to be regular, the metric functions must
satisfy the boundary conditions
\begin{equation}
f|_{r=r_{\rm H}}=m|_{r=r_{\rm H}}=n|_{r=r_{\rm H}}=0 \ ,
\ \omega|_{r=r_{\rm H}}=r_{\rm H} \Omega_{\rm H}  
\ , \label{bc4} \end{equation}
where $\Omega_{\rm H}$ is the horizon angular velocity, Eq.~(\ref{Omega}).
%
The gauge potential satisfies at the horizon the conditions (\ref{Phi})
\begin{equation}
\left. \zeta^\mu A_\mu \right|_{r=r_{\rm H}} =
\Phi_{\rm H} = \left. (a_0+\Omega_{\rm H} a_\varphi)\right|_{r=r_{\rm H}} \ , \ \ \
\left. \frac{d a_\varphi}{d r}\right|_{r=r_{\rm H}}=0
\ , \label{bc5} \end{equation}
with the constant horizon electrostatic potential $\Phi_{\rm H}$.

\subsection{Expansions}

From the
asymptotic expansion of the metric functions and the gauge field 
functions we can extract various parameters of the black hole:
\begin{eqnarray}
&&f \to 1 - \frac{M}{6\pi^2r^2} \ , \
m \to 1 - \frac{M}{12\pi^2r^2} \ , \
n \to 1 - \frac{M}{12\pi^2r^2} \ , \
\nonumber
\\
&&\omega \to \frac{J}{4\pi^2r^3} \ , \ 
a_0 \to \frac{Q}{4\pi^2r^2} \ , \ 
a_{\varphi} \to -\frac{\mu_{mag}}{4\pi^2r^2} \ ,
\end{eqnarray}
where $M$ is the mass, $J$ the angular momentum and $\mu_{mag}$ the magnetic
moment. From these charges we can calculate the black hole gyromagnetic ratio
$g = \frac{2M\mu_{mag}}{QJ}$.

Next we consider the expansion of the functions near the horizon. First we
present the expansion for non-extremal black holes. Once the previous
conditions are imposed, it can be seen that:
\begin{eqnarray}
f(r) &=& f_2 (r-r_{\rm H})^2 + o((r-r_{\rm H})^3), \nonumber \\
m(r) &=& m_2 (r-r_{\rm H})^2 + o((r-r_{\rm H})^3), \nonumber \\
n(r) &=& l_2 (r-r_{\rm H})^2 + o((r-r_{\rm H})^3), \\
\omega(r) &=& \Omega_{\rm H} + o(r-r_{\rm H}), \nonumber \\
a_0(r) &=& a_{0,0} + o((r-r_{\rm H})^2), \nonumber \\
a_{\varphi}(r) &=& a_{\varphi,0} + o((r-r_{\rm H})^2). \nonumber
\end{eqnarray}
These constants are implicitly related to three global charges ($M$, $Q$ and
$J$) in the non-extremal case.

In the case of extremal black holes, the radial dependence of the functions
in the vicinity of the horizon changes as follows:
\begin{eqnarray}
f(r) &=& f_4 r^4 + f_{\alpha} r^{({\alpha}+4)} + o(r^6), \nonumber \\
m(r) &=& m_2 r^2 + m_{\beta} r^{({\beta}+2)} + o(r^4), \nonumber \\
n(r) &=& l_2 r^2 + l_{\gamma} r^{({\gamma}+2)} + o(r^4), \nonumber \\
\omega(r) &=& \Omega_{\rm H} r + \omega_2 r^2 + o(r^3),  \\
a_0(r) &=& a_{0,0} + a_{0,\lambda} r^{\delta} + o(r^2), \nonumber \\
a_{\varphi}(r) &=& a_{\varphi,0} + l_{k,\mu} r^{\mu} + o(r^2).\nonumber
\end{eqnarray}
Since we are considering the extremal case, all these constants are implicitly
related to two global charges ($Q$ and $J$ for example). The coefficients
$\alpha$, $\beta$, $\gamma$, $\lambda$ and $\mu$ can be 
non-integer. In fact we have 
\begin{eqnarray}
&&0<\alpha<2 \ , \quad 0<\beta<2 \ ,\quad  0<\gamma<2 \ ,  \\
&&0<\delta<2 \ , \, \quad  0<\mu<2 \ ,  \quad  0<\nu<2 \ .\nonumber
\end{eqnarray}




In the pure Einstein-Maxwell case ($\lambda=0$) the expansion contains similarly non-integer exponents. This feature is found in both EM branches 
($i.e.$, in the MP branch and in the RN branch).

Thus the expansion at the horizon in general leads to
non-integer exponents. We observe that these exponents ($\alpha$, $\beta$,
$\gamma$, $\lambda$ and $\mu$) can become lower than
one. This could lead to a divergence of the first derivative of a
reparametrized function. Hence we have to be very careful with the
reparametrization we use for the numerics. We reparametrize the functions in the
following form:
\begin{eqnarray}
\hat m &=&  m(r) \nonumber
\\
\hat n &=&  n(r) \nonumber
\\
\hat f &=&  f(r)/x^2 \\
\hat \omega &=& (1-x)^{-2} \omega(r)
\nonumber \\
\hat a_{\varphi} &=& x^2 a_{\varphi}(r). \nonumber
\end{eqnarray}
We then integrate the functions $\hat m, \hat n, \hat f, \hat \omega$ and $\hat
a_{\varphi}$. The advantage of using these reparametrized functions is the following. Since the exponents ($\alpha$, $\beta$,
$\gamma$, $\lambda$ and $\mu$) are always greater than $0$, and the order of the
differential equations is lower or equal than $2$, there is no problem with
diverging functions or derivatives in the numerics.

Note, that all the redefined functions except for $\hat \omega$ now start with an
$x^2$-term in the compactified coordinate $x = r/(r+1)$. (The
reparametrization of $\omega$ is not related to the expansion at the horizon. It
is done in order to be able to fix the angular momentum by a boundary
condition). 





The boundary conditions must guarantee that the functions satisfy the
horizon expansion ($\hat m, \hat n, \hat f$ and $\hat a_{\varphi}$ start as
$x^2$). At the horizon, $x=0$, we can impose the angular velocity by making
$\hat \omega'(0)=\Omega_{\rm H}$. Note that here a prime indicates derivation with
respect to the compactified coordinate $x$. 

At $x=1$ we impose asymptotic flatness by requiring
\begin{eqnarray}
\hat m(1) &=& 1 \nonumber
\\
\hat n(1) &=& 1 
\nonumber\\
\hat f(1) &=& 1 \\
\hat \omega(1) &=& 0 
\nonumber\\
\hat a_{\varphi}(1) &=& 0.\nonumber
\end{eqnarray}
Alternatively, we can also impose the angular momentum by requiring $\hat
\omega'(1)=-J/(4\pi^2)$.

In the system of equations we are using, the $a_0$ function was
eliminated by introducing a first integral of the system in terms of the
electric charge. Using these boundary conditions we can obtain configurations
with selected pairs of values for $(Q, \Omega_{\rm H})$ or $(Q, J)$ for a given value
of $\lambda$.
Once a configuration is obtained, we can compute the global charges. 

From the asymptotic behaviour of the functions at $x=1$ we extract the mass
$M$, the angular momentum $J$, and the magnetic moment $\mu_{mag}$. With these
quantities we calculate the gyromagnetic factor $g$.
From the behaviour of the functions at the horizon $x=0$ we extract the
horizon mass $M_{\rm H}$, the horizon angular momentum $J_{\rm H}$, the horizon area $A_{\rm H}$
and the horizon electrostatic potential $\Phi_{H}$.
The Smarr formula is always satisfied within the precision of the black hole
configurations computed.

\section{Global solutions: Numerical results}

We now discuss the global EMCS black hole solutions obtained by numerical integration
as discribed above,
with emphasis on extremal configurations.
Here we focus on the most interesting case for the CS coupling constant $\lambda$,
namely $\lambda > 2$. We first discuss the relation between the near-horizon solutions
and the global solutions, and exhibit the intriguing branch structure of the latter.
Subsequently we address the sequence of $J=0$ solutions and reveal the node
structure of the two functions $a_\varphi(r)$ and $\omega(r)$.
Finally, we discuss the domain of existence of the global solutions.

Concerning the lower values of $\lambda$, let us remark that for
$0<\lambda < 1.91$ the global extremal solutions are in one to one correspondence
to the near-horizon solutions, while in the transitions region $1.91 < \lambda <2$
this does no longer seem to be the case.
When approaching the RN solution for values of $\lambda$ in the transition region
and positive values of the charge,
the functions of the rotating global extremal solutions  
develop highly divergent derivatives.
This forbids definite statements on the existence and properties
of the numerically constructed global solutions in this area of parameter space.
However, our analysis indicates, that the rotating global extremal solutions are disconnected
from the static extremal RN black hole in the transition region.
In particular, we did not obtain any non-static extremal $J=0$ solutions 
for $\lambda < 2$.

\subsection{Near-horizon versus global solutions: the branch structure of extremal black holes}

\begin{figure}[t!]
\begin{center}
\mbox{\hspace*{-1.0cm}
\subfigure[][]{
\includegraphics[height=.4\textheight, angle
=270]{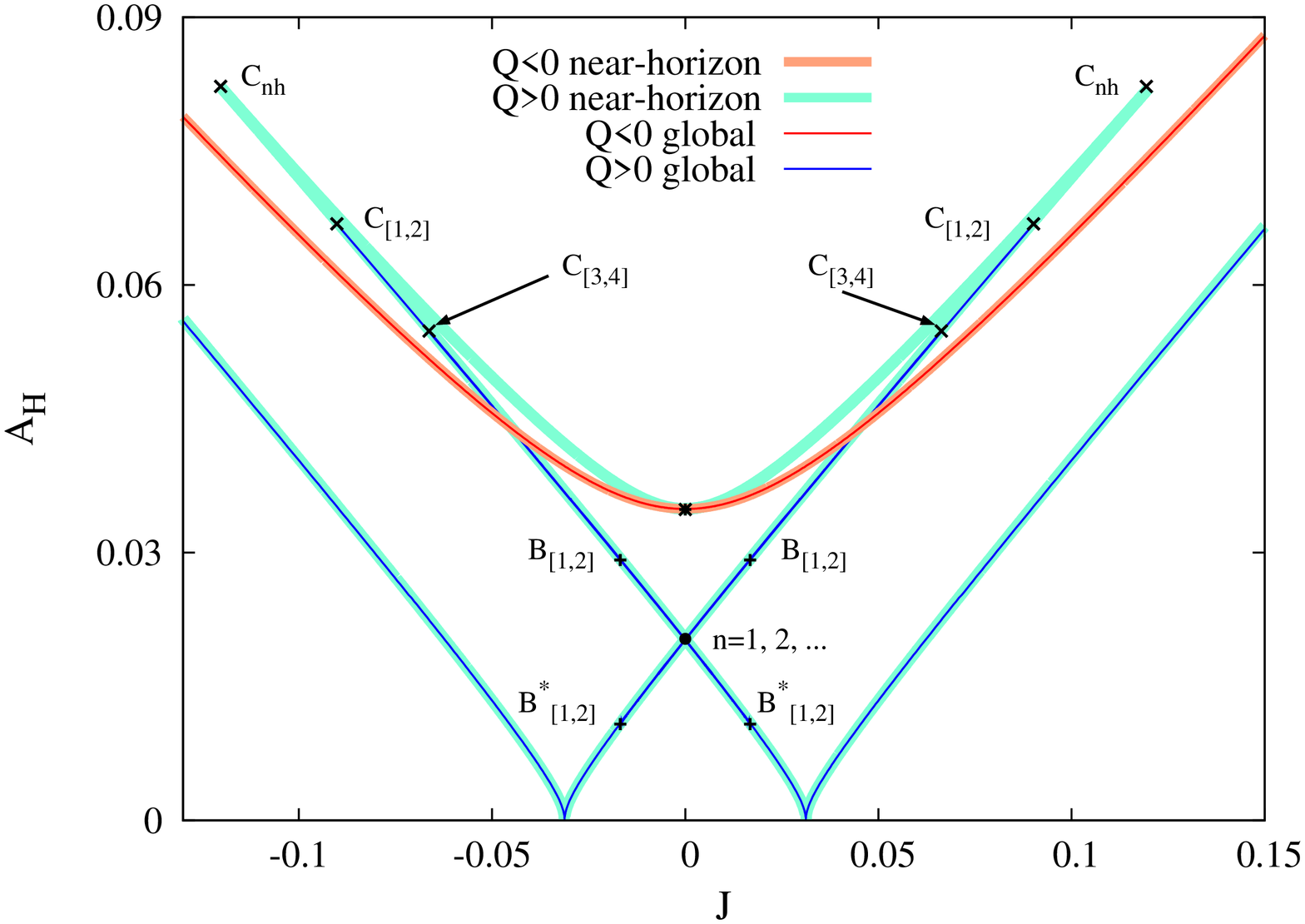}
\label{plot_J_Ah_lambda_5_extremal_long_paper.ps}
}
\hspace*{-0.6cm}
\subfigure[][]{
\includegraphics[height=.4\textheight,
angle=270]{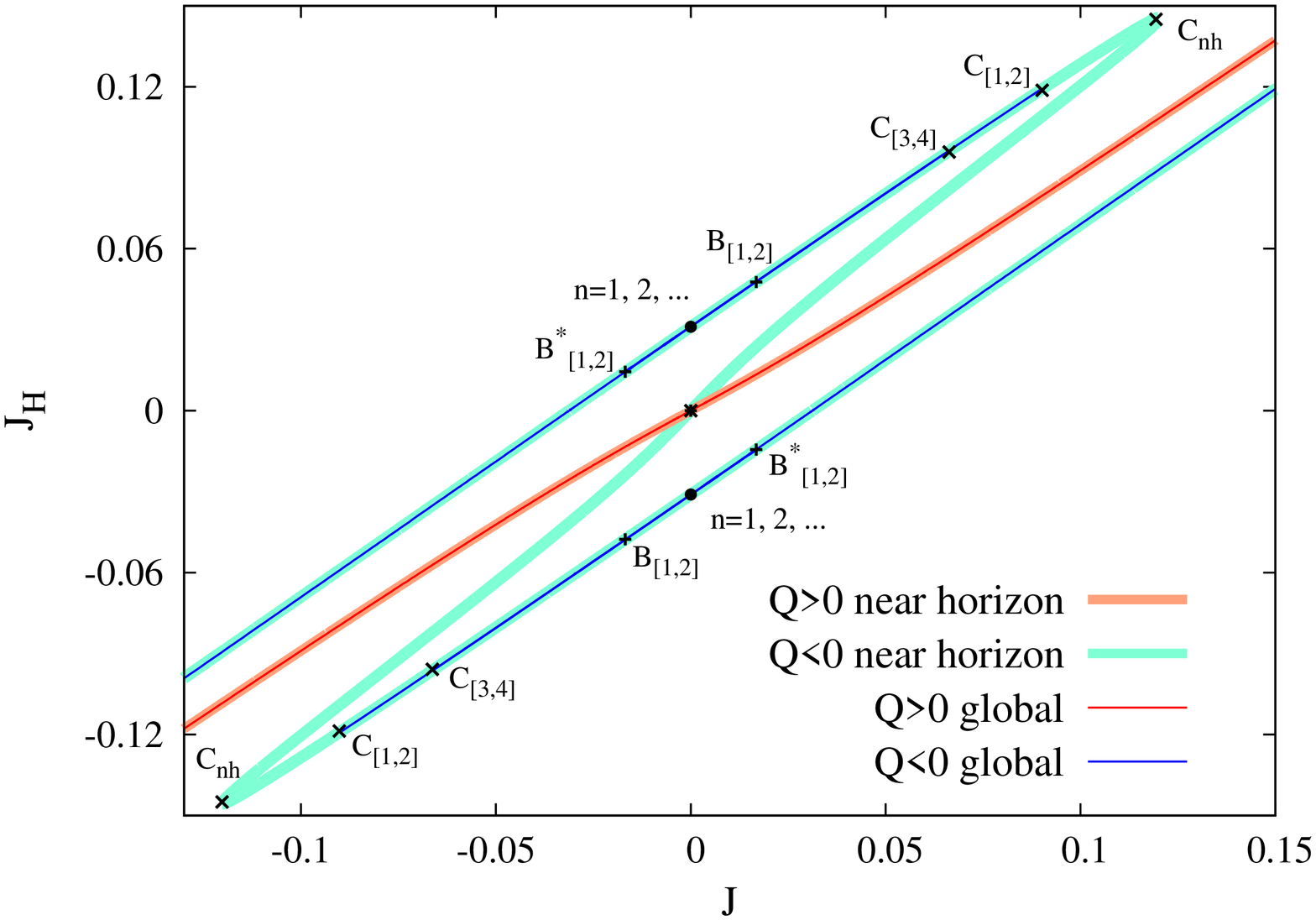}
\label{plot_J_Jh_lambda_5_extremal_long_paper.ps}
}
}
\end{center}
\caption{\small
Global versus near-horizon solutions:
The horizon area $A_{\rm H}$ (a)  and the horizon angular momentum $J_{\rm H}$
versus the angular momentum $J$ for charge $|Q|=1$
and for CS coupling $\lambda=5$. 
The thick lines represent the near horizon solutions
for positive charge (green) and negative charge (red), 
while the thin lines the global solutions
for positive charge (blue) and negative charge (red).
The cusps are marked by $\times$ and denoted by $C_{[...]}$,  
the bifurcation points are marked by $+$ and denoted by $B_{[...]}$ (see text).
The black asterisk represents the extremal RN solution.
}
\label{fig4}
\end{figure}

Let us now recall the near-horizon solutions and overlay the corresponding
properties of the global solutions for direct comparison.
In Fig.~\ref{fig4}
we exhibit the horizon area $A_{\rm H}$ and the horizon angular momentum $J_{\rm H}$
versus the total angular momentum $J$ for both sets of solutions.
For definiteness, we have chosen CS coupling $\lambda_{\rm CS}=5$
and charge $Q=\pm 1$.
However, the pattern observed has the same structure for all values of 
the CS coupling constant $\lambda > 2$
and values of the charge $Q$.
The near-horizon solutions are marked by thick lines, while
the global solutions are marked by thin lines in the figures.

Clearly, for the negative charge solutions, shown in red, 
both global and near-horizon solutions match.
Thus in this case all near-horizon solutions are realized globally.
However, for positive charge solutions,
shown in green (near-horizon) and blue (global), the situation is very different.
The branches of global solutions end in the cusps $C_{[1,2]}$,
while the near-horizon solution branches extend further to $C_{\rm nh}$.
Thus the near-horizon solutions between both cusps are not realized globally. 
Moreover, the full near-horizon branch connecting
the cusp $C_{\rm nh}$ and the extremal RN solution, which is marked
by an asterisk, is neither realized globally --
with the exception of the RN solution itself, of course.
 
Therefore we conclude, that the static extremal RN solution with $Q>0$ is
rotationally isolated from the family of global stationary extremal solutions, 
when $\lambda>2$ (and presumably already for $\lambda > 1.91$).
The horizon area $A_{\rm H}$ of the global solutions with $J=0$
is significantly smaller than the horizon area of the RN solution.
Moreover, they carry finite horizon angular momentum $J_{\rm H}$.

We note, that the existence of near-horizon solutions,
which do not possess global counterparts, has been noted first
by Chen et al.~\cite{Chen:2008hk}
for the extremal dyonic black holes of $D=4$ Gau\ss -Bonnet gravity.
However, in the EMCS case the relation between near-horizon solutions
and global solutions is even more surprising, as discussed in the following.
By considering the physical quantities, not accessible in the
near-horizon formalism, like the mass $M$ or the
horizon angular velocity $\Omega_{\rm H}$,
we now present the intricate pattern of branches 
of the global rotating extremal black holes.

\begin{figure}[h!]
\begin{center}
\mbox{\hspace*{-1.0cm}
\subfigure[][]{
\includegraphics[height=.4\textheight, angle
=270]{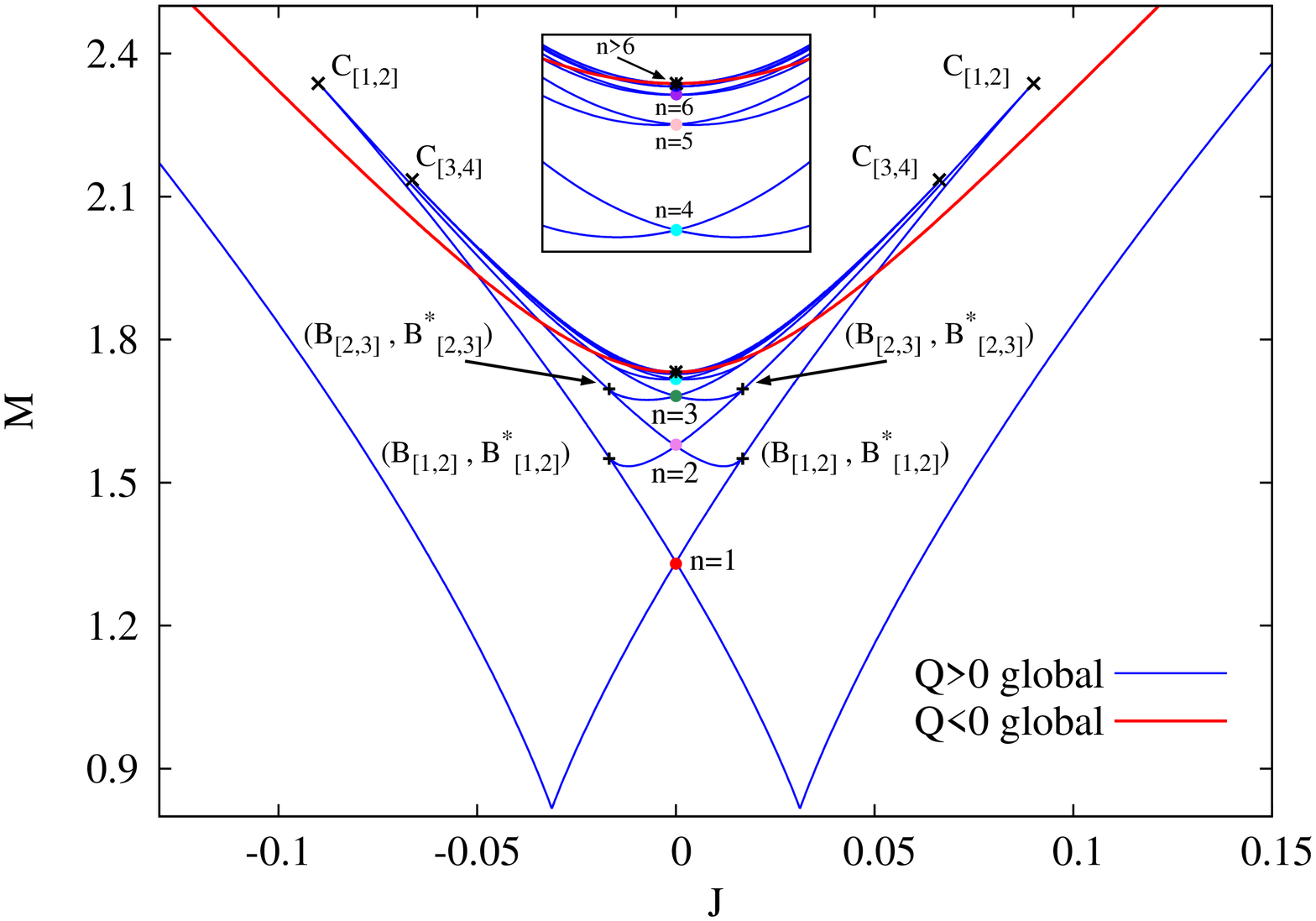}
\label{plot_J_M_lambda_5_extremal2_long_paper.ps}
}
\hspace*{-0.6cm}
\subfigure[][]{
\includegraphics[height=.4\textheight,
angle=270]{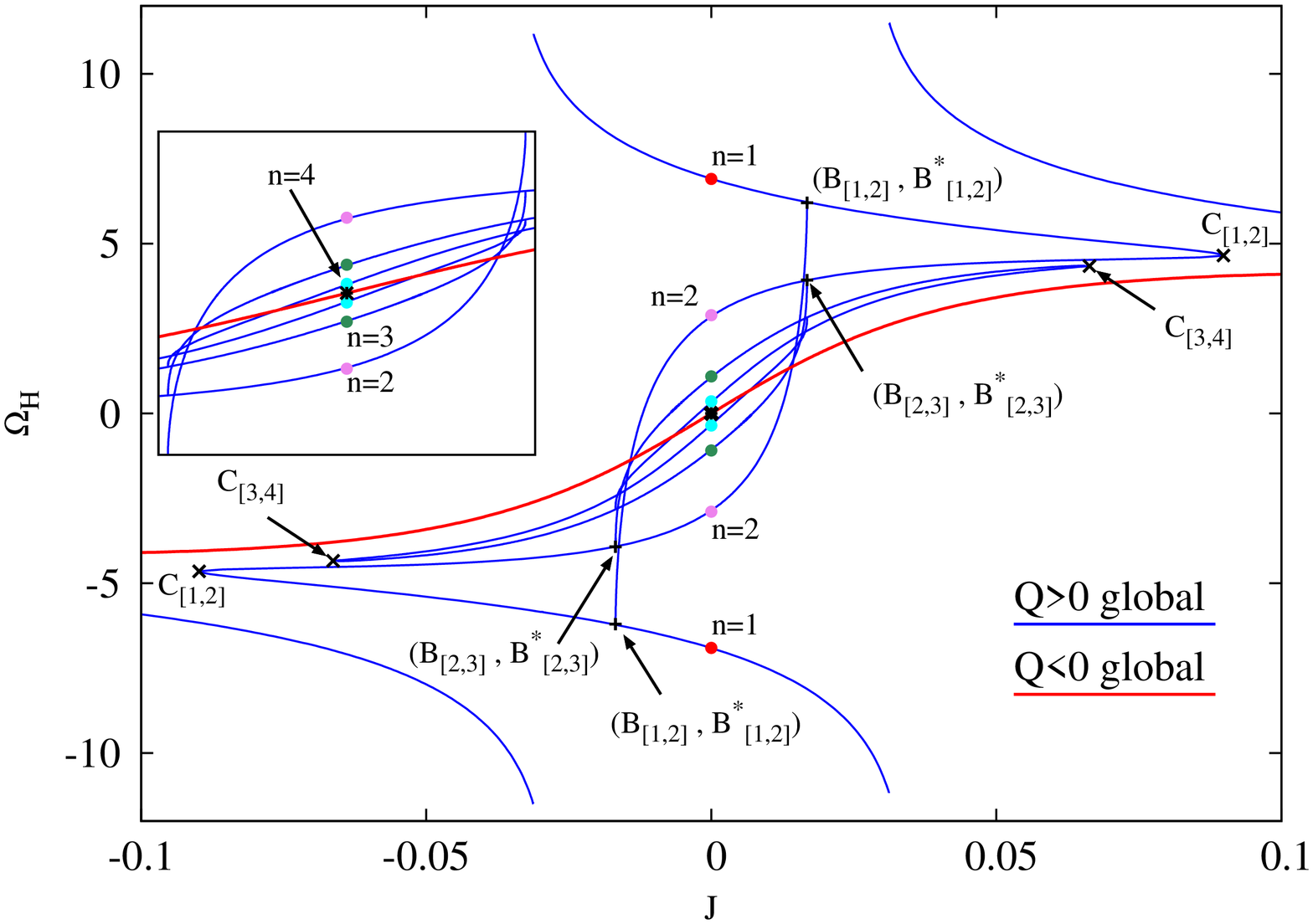} 
\label{plot_J_Omh_lambda_5_extremal_long_paper.ps}
}
}
\end{center}
\vspace*{-0.5cm}
\caption{\small
Global solutions:
The total mass $M$ (a) and the horizon angular velocity $\Omega_{\rm H}$ (b)
versus the angular momentum $J$ for charge $Q=1$ (blue), $Q=-1$ (red)
and for CS coupling $\lambda=5$. 
The cusps are marked by $\times$ and denoted by $C_{[...]}$,
the bifurcation points are marked by $+$ and denoted by $B_{[...]}$ (see text).
The non-static $J=0$ solutions are numbered by $n$ and marked
by red dots in different colors.
The black asterisk represents the extremal RN solution.
}
\label{fig5}
\end{figure}

We exhibit in Fig.~\ref{fig5} the mass $M$ (a) and the 
horizon angular velocity $\Omega_{\rm H}$ (b) versus the total angular momentum $J$
of the global extremal black hole solutions.
We have chosen CS coupling constant $\lambda=5$ 
and electric charge $Q=\pm 1$.
These figures contrast the simple branch structure for negative charge
(red) with the highly complicated branch structure for positive charge (blue).

\begin{figure}[h!]
\begin{center}
\mbox{
\includegraphics[height=.5\textheight,
 angle
=270]{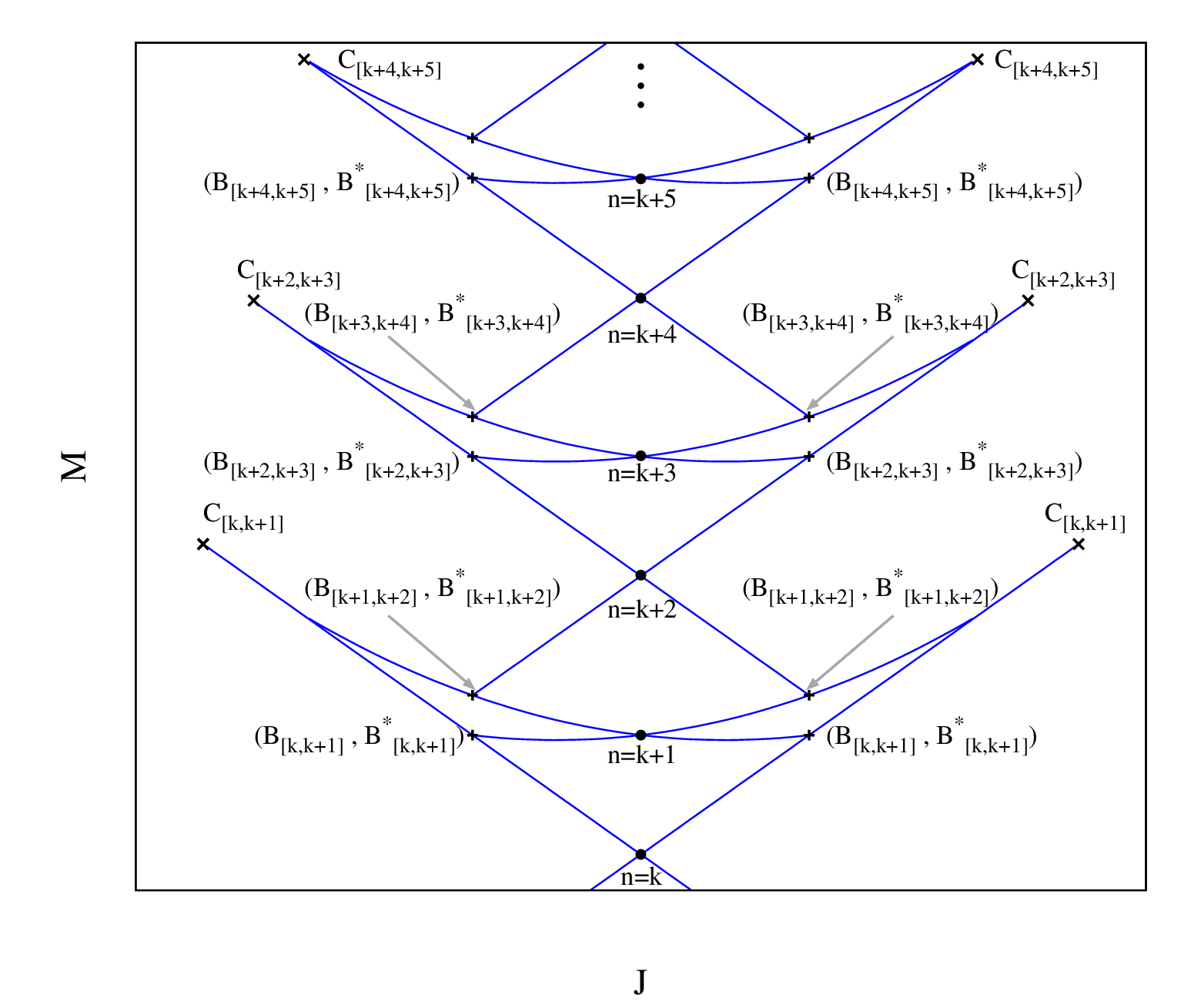}
}
\end{center}
\caption{\small
Schematic representation of the branch structure 
of the global solutions:
The total mass $M$ versus the angular momentum $J$.
The cusps are marked by $\times$ and denoted by $C_{[...]}$,
the bifurcation points are marked by $+$ and denoted by $B_{[...]}$.
The non-static $J=0$ solutions are numbered by $n$.
}
\label{fig6}
\end{figure}

Let us begin the detailed discussion of the branch structure
for positive $Q$
with Fig.~\ref{plot_J_M_lambda_5_extremal2_long_paper.ps},
which has a number of similar features to
Fig.~\ref{plot_J_Ah_lambda_5_extremal_long_paper.ps}.
For instance, the two points in
Fig.~\ref{plot_J_M_lambda_5_extremal2_long_paper.ps},
where a minimal value of the mass is encountered, 
correspond to the two singular solutions
in Fig.~\ref{plot_J_Ah_lambda_5_extremal_long_paper.ps},
which have vanishing horizon area.
We now choose one of these particular solutions, 
the one with $J<0$,
as the starting point of our analysis of the branch structure.

When we decrease the angular momentum from this $J_{\rm H}<0$, $A_{\rm H}=0$ solution, 
we obtain a single infinite branch of extremal solutions, 
whose mass increases monotonically with decreasing $J$. 
In contrast, when we increase the angular momentum from 
this $J_{\rm H}<0$, $A_{\rm H}=0$ solution, 
we cross the first non-static $J=0$ solution,
which we label $n=1$. 
Increasing $J$ further, 
we then cross a particular solution,
which we label $B_{[1,2]}$. 
Eventually, this branch ends at a finite maximal value of $J$,
where a cusp is encountered, 
which we label $C_{[1,2]}$.
The solution at the cusp is perfectly regular.

From the cusp $C_{[1,2]}$ a new branch of solutions bends backwards
towards smaller masses, as the angular momentum is decreased.
Along this branch first another particular solution is encountered,
which we label $B_{[2,3]}$.
Then the branch crosses the next non-static $J=0$ solution, 
labeled $n=2$. 
It eventually ends in a solution labeled $B^*_{[1,2]}$,
when it reaches the $J \to - J$ symmetric first branch
at the solution $B_{[1,2]}$.
Thus, we have encountered a bifurcation point, 
where the two branches meet,
which has two distinct solutions,
$B^*_{[1,2]}$ and $B_{[1,2]}$,
that possess the same global charges.

This reveals the presence of non-uniqueness among the extremal solutions.
While the global charges of these solutions are the same, 
and their horizon angular velocities are also the same,
as seen in  Fig.~\ref{plot_J_Omh_lambda_5_extremal_long_paper.ps},
they differ in their horizon area, 
as seen in Fig.~\ref{plot_J_Ah_lambda_5_extremal_long_paper.ps}.
The area of $B^*_{[1,2]}$ is considerably smaller than the
area of $B_{[1,2]}$.

These two sets of symmetric branches with cusps $C_{[1,2]}$
and bifurcation points $B_{[1,2]}$, $B^*_{[1,2]}$
form only the lowest mass solutions of a whole tower
of branches, built in an analogous ever repeating manner.
We have already noted the presence of the particular
(symmetric) points $B_{[2,3]}$. 
Here the next (symmetric) bifurcations are encountered,
where new branches arise, whose initial solutions
are labeled $B^*_{[2,3]}$.

Starting at the solution $B^*_{[2,3]}$ with negative $J$,
we cross the next non-static $J=0$ solution, $n=3$.
Increasing $J$ further,
we then cross the solution at the next bifurcation point
which we label $B_{[3,4]}$,
and then encounter the next cusp $C_{[3,4]}$.
From $C_{[3,4]}$ a new branch of solutions bends backwards
towards smaller masses, as the angular momentum is decreased.
Now the next bifurcation point is encountered,
$B_{[4,5]}$, then the next non-static $J=0$ solution, $n=4$, is passed,
and the branch ends at the bifurcation point,
formed by $B^*_{[3,4]}$ and $B_{[3,4]}$.

This general scheme is exhibited schematically in Fig.~\ref{fig6}.
Starting from the solution $B^*_{[k+1,k+2]}$ with negative $J$,
first the $(k+2)$th non-static $J=0$ solution is encountered,
then the bifurcation point $B_{[k+2,k+3]}$,
and subsequently the cusp $C_{[k+2,k+3]}$.
From $C_{[k+2,k+3]}$ a new branch bends backwards
towards smaller masses, encountering first
the bifurcation point $B_{[k+3,k+4]}$
then the $(k+3)$th non-static $J=0$ solution,
and it ends at the bifurcation point, 
formed by $B^*_{[k+2,k+3]}$ and $B_{[k+2,k+3]}$.

For any $k$,
the two solutions at the bifurcation points 
$B_{[k+2,k+3]}$ and $B^*_{[k+2,k+3]}$
possess the same global charges,
whereas their horizon area and horizon angular momentum differ.
Thus these solutions constitute a sequence of extremal solutions,
that violate uniqueness.
The horizon angular velocities of the solutions
$B_{[k+2,k+3]}$ and $B^*_{[k+2,k+3]}$, however, possess the same value.
This is clearly seen for the lowest $k$ in 
Fig.~\ref{plot_J_Omh_lambda_5_extremal_long_paper.ps}.
This figure also highlights the singular nature of the
points with vanishing area,
showing that the horizon angular velocity 
jumps from $\Omega_{\rm H}$ to $-\Omega_{\rm H}$ at these points.

Having generated in this way a whole sequence of branches,
labeled by $n$,
let us inspect the extent of these branches.
Here we observe that with increasing $n$,
the cusps $C_{n,n+1}$ occur at decreasing values of $|J|$.
Thus the range of extremal global solutions for fixed $|J|$
decreases with increasing $n$. 

Coming finally back to the comparison of global and near-horizon solutions,
we conclude that {\it a given near-horizon solution
can correspond to} 
\begin{itemize}
\itemsep=0pt
\item[i)]
{\it more than one global solution},
\item[ii)]
{\it  precisely one global solution}, 
\item[iii)]
{\it no global solution at all}.
\end{itemize}
In fact, we conjecture, that a given near-horizon solution
may even correspond to an infinite set of global solutions,
as discussed in the following.

\boldmath
\subsection{Radial excitations: $J=0$ solutions}
\unboldmath

Let us now address the question of how these branches of global
extremal solutions differ. In particular,
we would like to associate the integer $n$ counting the
branches with a physical property of the
solutions along these branches.
To that end, let us first focus on the set of $J=0$ solutions.
As seen in Fig.~\ref{plot_J_M_lambda_5_extremal2_long_paper.ps},
the mass $M_n$ of each pair of degenerate $J=0$ solutions
increases with $n$. This suggests, that the solutions
form a sequence of higher and higher excited states,
with the lowest mass solution forming the ground state
or fundamental solution.

Excited states of otherwise very similar solutions are 
typically found, when radial excitations are allowed.
Examples range from the hydrogen atom, 
via sphaleron solutions in flat space
\cite{Kunz:1988sx}, 
to hairy black holes \cite{Volkov:1990sva},
all featuring an infinite number of radial excitations of the ground state.

\begin{figure}[t!]
\begin{center}
\mbox{\hspace*{-0.8cm}
\subfigure[][]{
 \includegraphics[height=.4\textheight,
 angle=270]{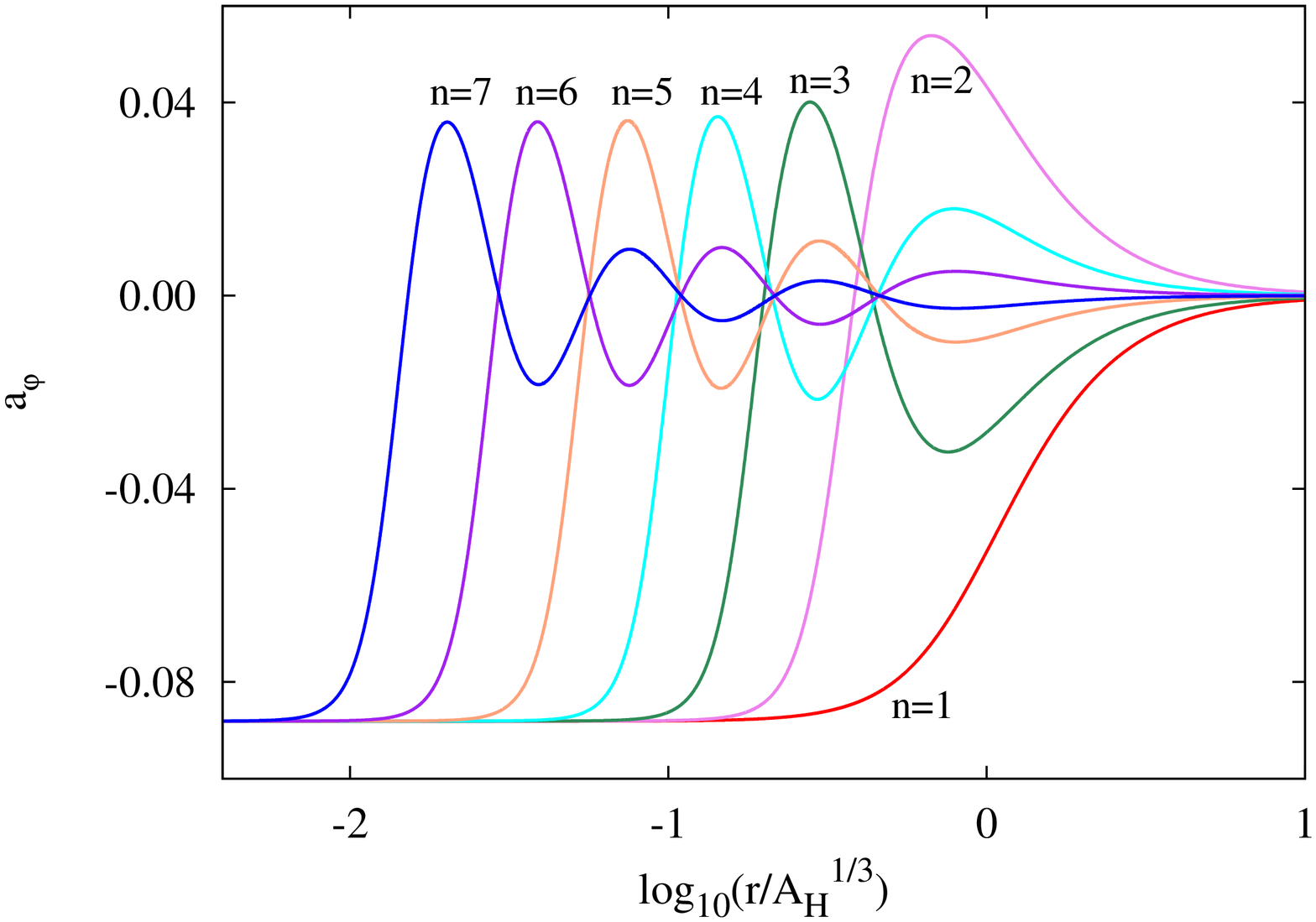}
 \label{plot_x_ak_lambda_5_J=0_log3_new2_long_paper.ps}
 }
\hspace{-0.8cm}
\subfigure[][]{
\includegraphics[height=.4\textheight,
 angle
=270]{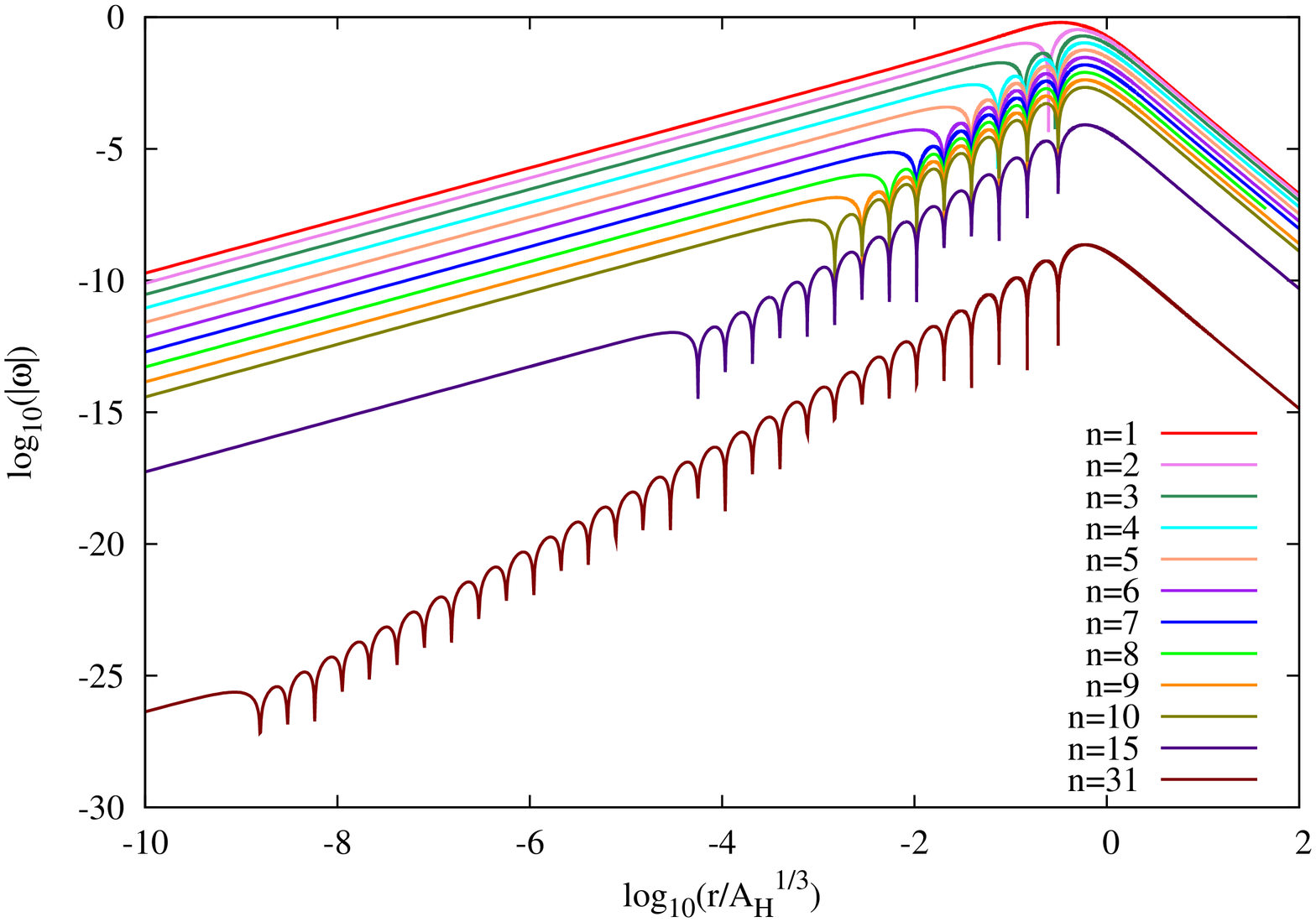}
\label{plot_x_omega_lambda_5_J=0_log3_long_paper.ps}
}
}
\end{center}
\caption{\small
Global solutions:
The gauge field function $a_\vphi$ (a) 
and the logarithm of the modulus of the  metric function $\omega$ (b)
versus the logarithm of the scaled radial coordinate
$\log_{10}\left(r/A_{\rm H}^{1/3} \right)$.
Shown are the lowest radial excitations
$n=1,...,7$ of $a_\vphi$ (a) and $n=1,...,10$, 15 and 31 
of $\omega$ (b) for charge $Q=1$ and CS coupling constant $\lambda=5$.
}
\label{fig7}
\end{figure}

Indeed, it is the occurrence of radial excitations,
which results in the sequence of $J=0$ solutions
and the associated intriguing pattern of branches
of the global extremal black holes.
We exhibit in Fig.~\ref{fig7} the gauge field function
$a_\varphi$ (a) and the metric function $\omega$ (b)
versus the radial coordinate.
Again we have chosen charge $Q=1$ and CS coupling constant
$\lambda=5$.

As the radial coordinate in the figure we have
employed the logarithm of the scaled radial coordinate
$\log_{10}\left(r/A_{\rm H}^{1/3} \right)$.
Recall, that all these solutions possess the same
horizon area $A_{\rm H}$, 
as seen in Fig.~\ref{plot_J_Ah_lambda_5_extremal_long_paper.ps}.
The first node always refers to spatial infinity.
We note, that in this logarithmic scale,
the nodes of the solutions are located at roughly equal spacings.
Moreover, the $k$th node of the excited solutions with more 
than $k$ nodes are located very close to each other.
(This is also known for the hydrogen atom and other
radial excitations.)

For the metric function $\omega$ we have chosen to
exhibit the logarithm of its modulus, since in contrast to $a_{\varphi}$
it varies over many orders of magnitude, as $n$ is increased.
However, as seen in Fig.~\ref{plot_x_omega_lambda_5_J=0_log3_long_paper.ps},
beyond the nodes, for small values of the radial coordinate,
the functions $\log_{10}|\omega|$ exhibit the same slope for all $n$.
This also holds for large values of the radial coordinate.
Besides the ten lowest $n$,
Fig.~\ref{plot_x_omega_lambda_5_J=0_log3_long_paper.ps} also exhibits
the metric function for $n=15$ and $n=31$.
This clearly shows, that the pattern continues to high values of $n$.

Solutions with $n$ in the thirties are the solutions 
with the highest node numbers we could achieve numerically so far.
However, we conjecture, that this sequence of $J=0$ solutions
can be continued to arbitrarily high values of $n$.
Thus we conjecture, that we have obtained the lowest members
of an {\sl infinite sequence} of radially excited $J=0$ solutions.
If this is really the case, then the corresponding near-horizon
solutions will correspond to an infinite sequence of global solutions.

As mentioned already, with increasing node number $n$,
the mass $M_n$ increases monotonically. In fact, it
converges monotonically from below to the mass of the extremal
static RN black hole, $M_{\rm RN}$. 
The Smarr formula then implies, 
that the horizon electrostatic potential $\Phi_n$
converges likewise to the extremal static RN value, $\Phi_{\rm RN}$.
At the same time, the horizon angular velocity $\Omega_n$
converges to zero, and thus also to the extremal static RN value.
The horizon area $A_{\rm H}$, however, is independent
of $n$ and has the same value for all the members
of the (presumably) infinite sequence, and differs
significantly from the extremal static RN value.
The same holds true for the
magnitude of the horizon angular momentum $J_{\rm H}$.


Let us address now the $\lambda$-dependence of the observed
pattern of extremal $J=0$ solutions.
As pointed out already, there is
a lower critical value $\lambda_{\rm cr}$,
beyond which no rotating $J=0$ solutions exist.
When this critical value is approached from above,
the mass of the rotating $J=0$ solutions should 
tend to the mass of the extremal static RN solution,
in order to have a smooth emergence of these solutions
at the critical value.

\begin{figure}[t!]
\begin{center}
\mbox{\hspace*{-1.0cm}
\includegraphics[height=.4\textheight, angle
=270]{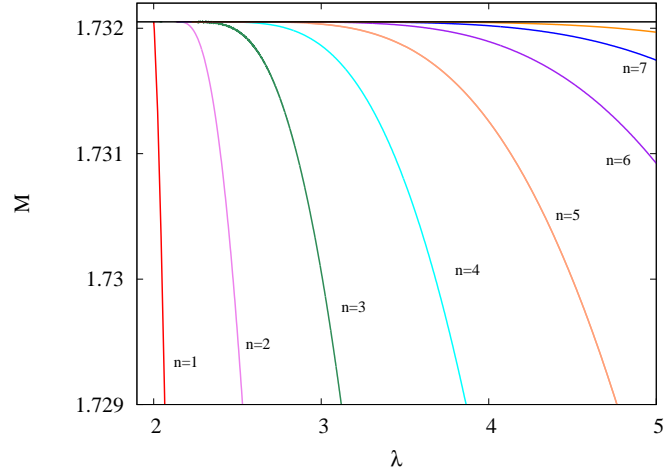}
}
\end{center}
\vspace*{-0.5cm}
\caption{\small
Global solutions:
The total mass $M$ versus the CS coupling constant $\lambda$.
Shown are the lowest radial excitations $n=1,...,7$
for charge $Q=1$.
}
\label{fig8}
\end{figure}

In Fig.~\ref{fig8}
we demonstrate that this is indeed the case.
Here we present the mass $M$ of the rotating $J=0$ solutions
with charge $Q=1$ and up to seven nodes
versus the CS coupling $\lambda$.
As expected the mass $M_n$ of the set of solutions with $n$ nodes
approaches the mass of the static extremal RN solution $M_{\rm RN}$,
when the CS coupling tends to $\lambda=2$.
In particular, we observe that
the larger the value of $n$, the closer is the mass $M_n$
to the limiting mass $M_{\rm RN}$ for fixed $\lambda$.
To discern the branch structure, it is therefore preferable
to employ values of $\lambda$ not too close to the critical value.
This explains our choice $\lambda =5$ in most figures.

\begin{figure}[t!]
\begin{center}
\mbox{
\subfigure[][]{\hspace{-0.8cm}
\includegraphics[height=.4\textheight,
 angle
=270]{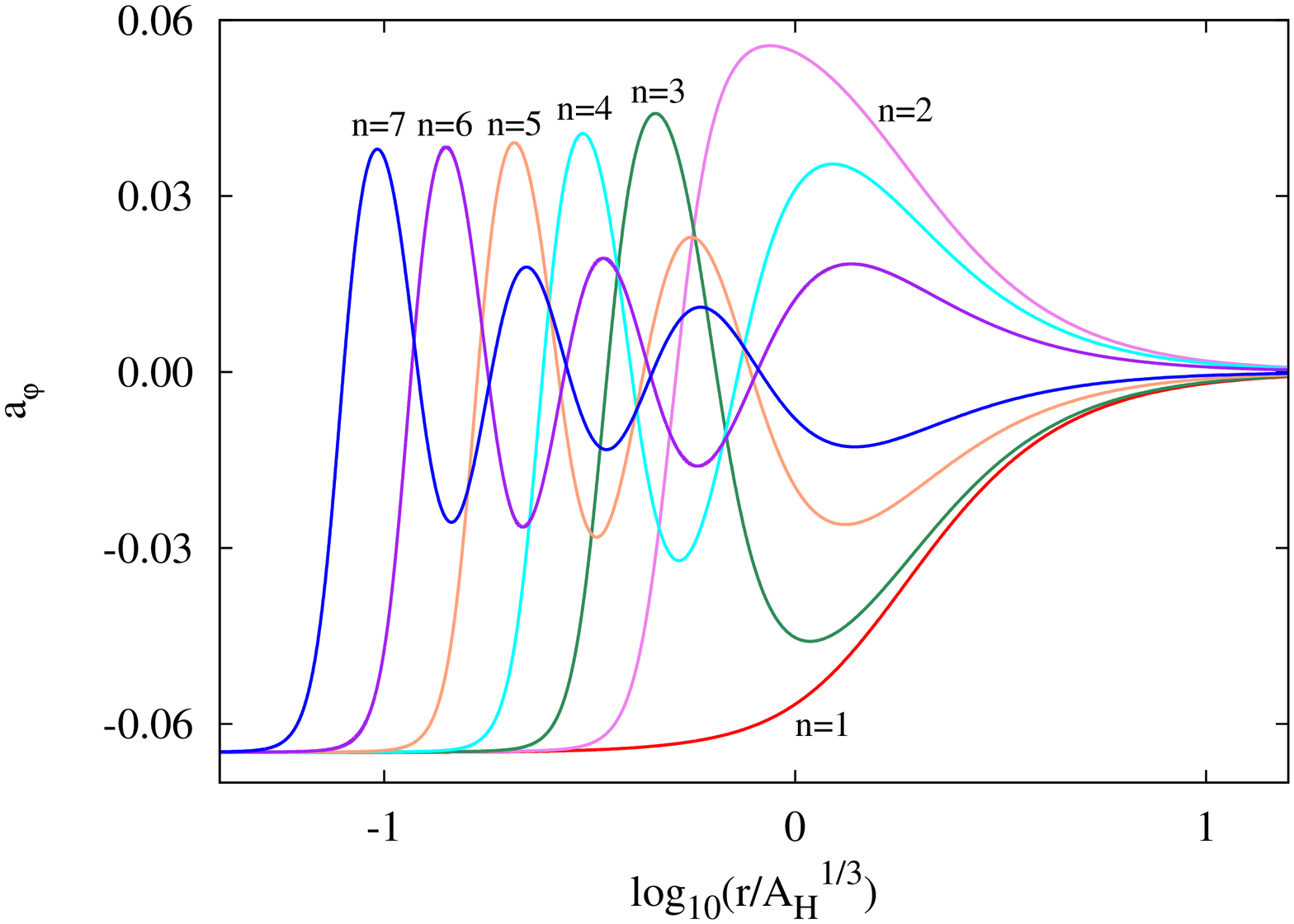}
\label{plot_x_ak_lambda_10_J=0_log3_new2_long_paper.ps}
}
\hspace{-0.8cm}
\subfigure[][]{
\includegraphics[height=.4\textheight,
 angle
=270]{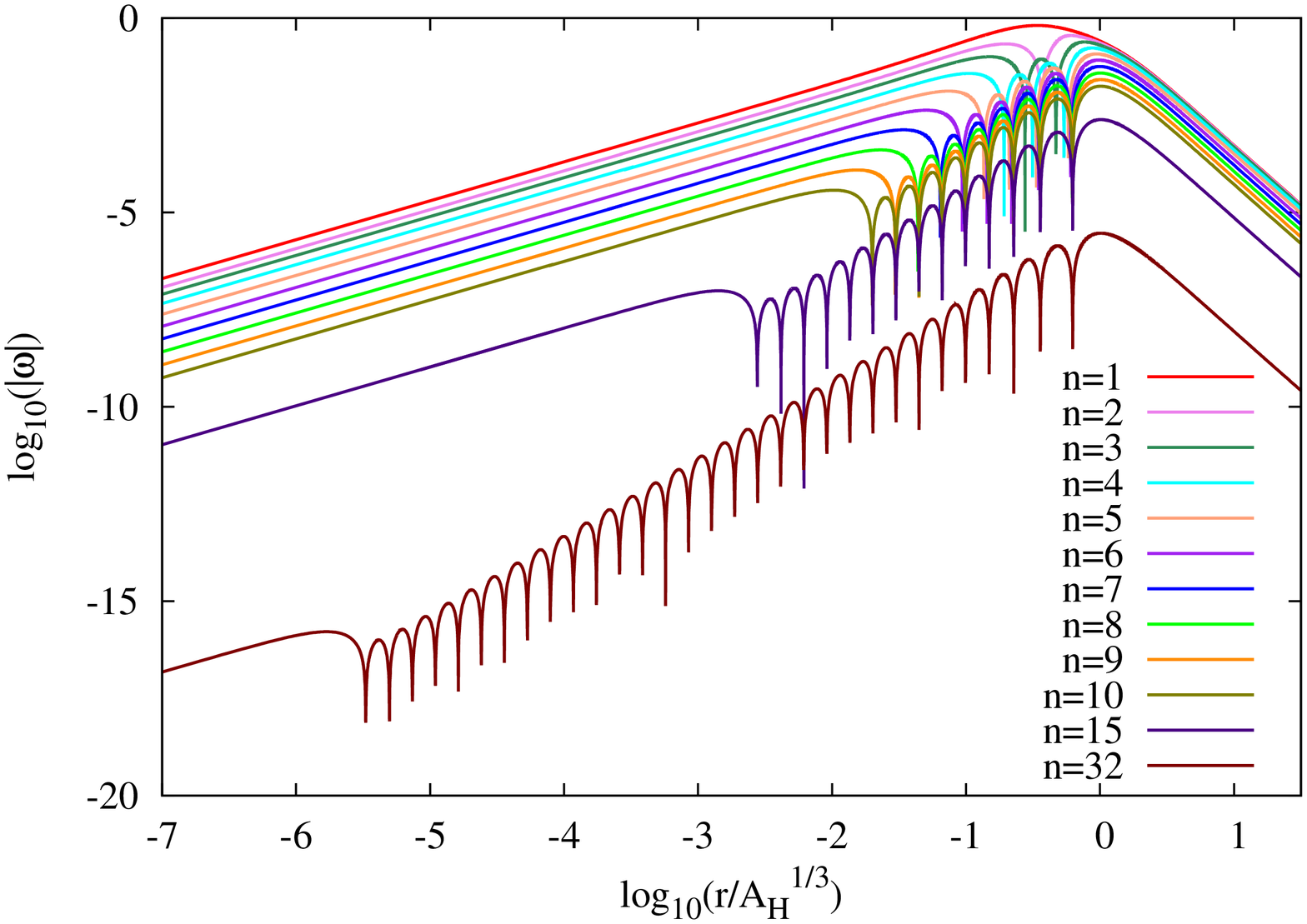}
\label{plot_x_omega_lambda_10_J=0_log3_long_paper.ps}
}
}
\end{center}
\caption{\small
Global solutions:
The gauge field function $a_\vphi$ (a)
and the logarithm of the modulus of the  metric function $\omega$ (b)
versus the logarithm of the scaled radial coordinate
$\log_{10}\left(r/A_{\rm H}^{1/3} \right)$.
Shown are the lowest radial excitations
$n=1,...,7$ of $a_\vphi$ (a) and $n=1,...,10$, 15 and 32
of $\omega$ (b) for charge $Q=1$ and CS coupling constant $\lambda=10$.
}
\label{fig9}
\end{figure}

In Fig.~\ref{fig9} we demonstrate the sequence of $J=0$ solutions
for a higher value of the CS coupling constant, $\lambda=10$.
Obviously, the figures for the gauge function and the
metric function are completely analogous to their counterparts,
exhibited in Fig.~\ref{fig7}.

The nodal excitations seen in the sequence of $J=0$ solutions,
are also present in the general set of global extremal solutions.
Along the various branches of solutions, new nodes appear from 
or disappear towards radial infinity.
Thus all global extremal solution possess an associated node number.
However, the extent of the branches decreases with increasing $n$.
Thus for a given $J \ne 0$, only a finite number of radial
excitations may exist.

Likewise, also the non-extremal solutions possess 
an associated node number. Fixing, for instance, the 
horizon radius, families of non-extremal solutions are obtained,
which exhibit an intricate branch structure reminiscent of
the one of the extremal solutions, but with less branches.
However, the number of branches increases the more, the closer the
family of non-extremal solutions approaches the 
set of extremal solutions.

Finally let us remark that
the node number might possibly be considered in the quest to regain
uniqueness.

\boldmath
\subsection{Domain of existence for $\lambda>2$}
\unboldmath

\begin{figure}[h!]
\begin{center}
\mbox{\hspace{-1.5cm}
\subfigure[][]{
\includegraphics[height=.35\textheight, angle =270]{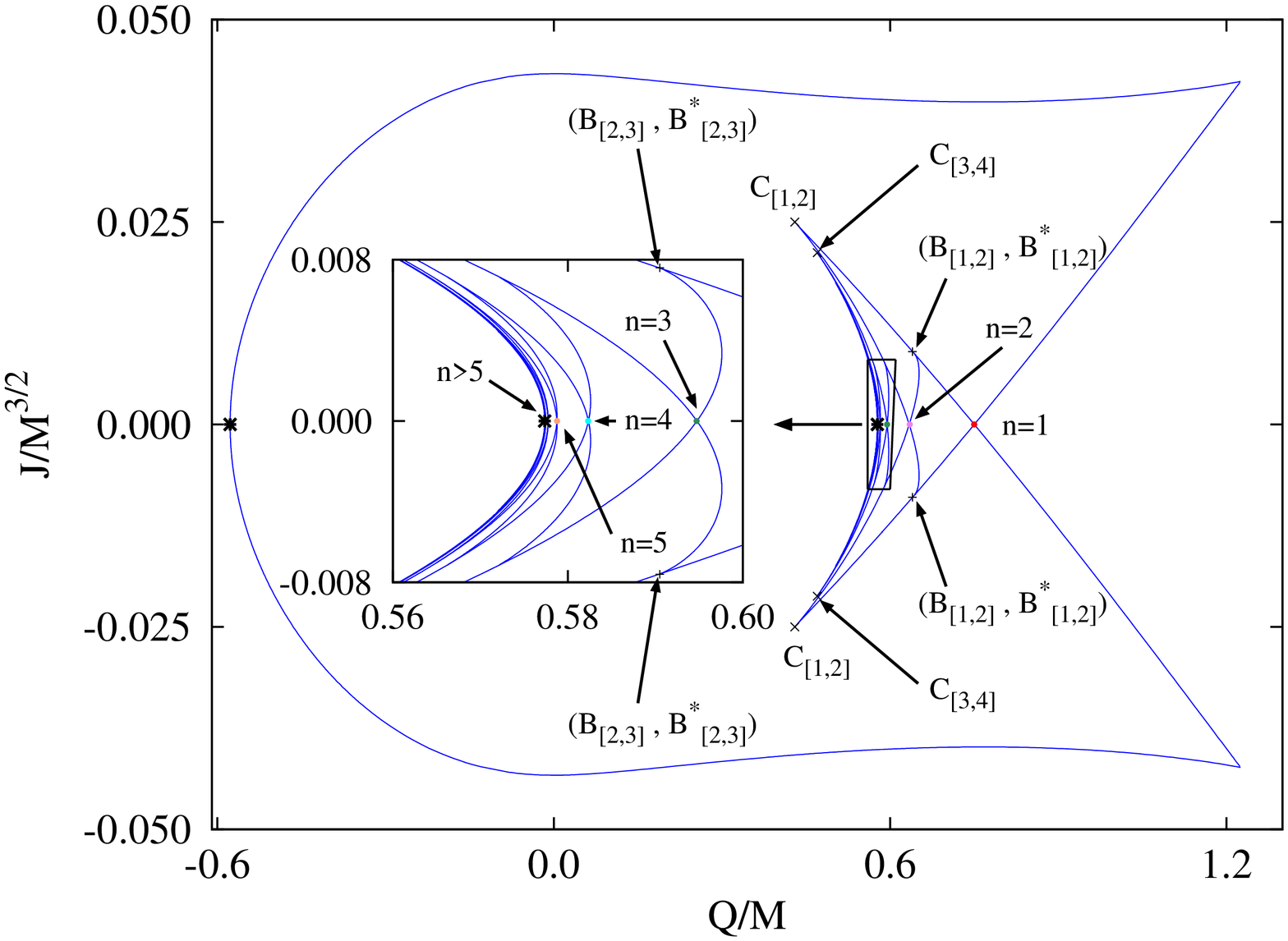}
\label{plot_q_j_lambda_5_extremal_z2a_long_paper.ps}
}
\subfigure[][]{\hspace{-0.5cm}
\includegraphics[height=.35\textheight, angle =270]{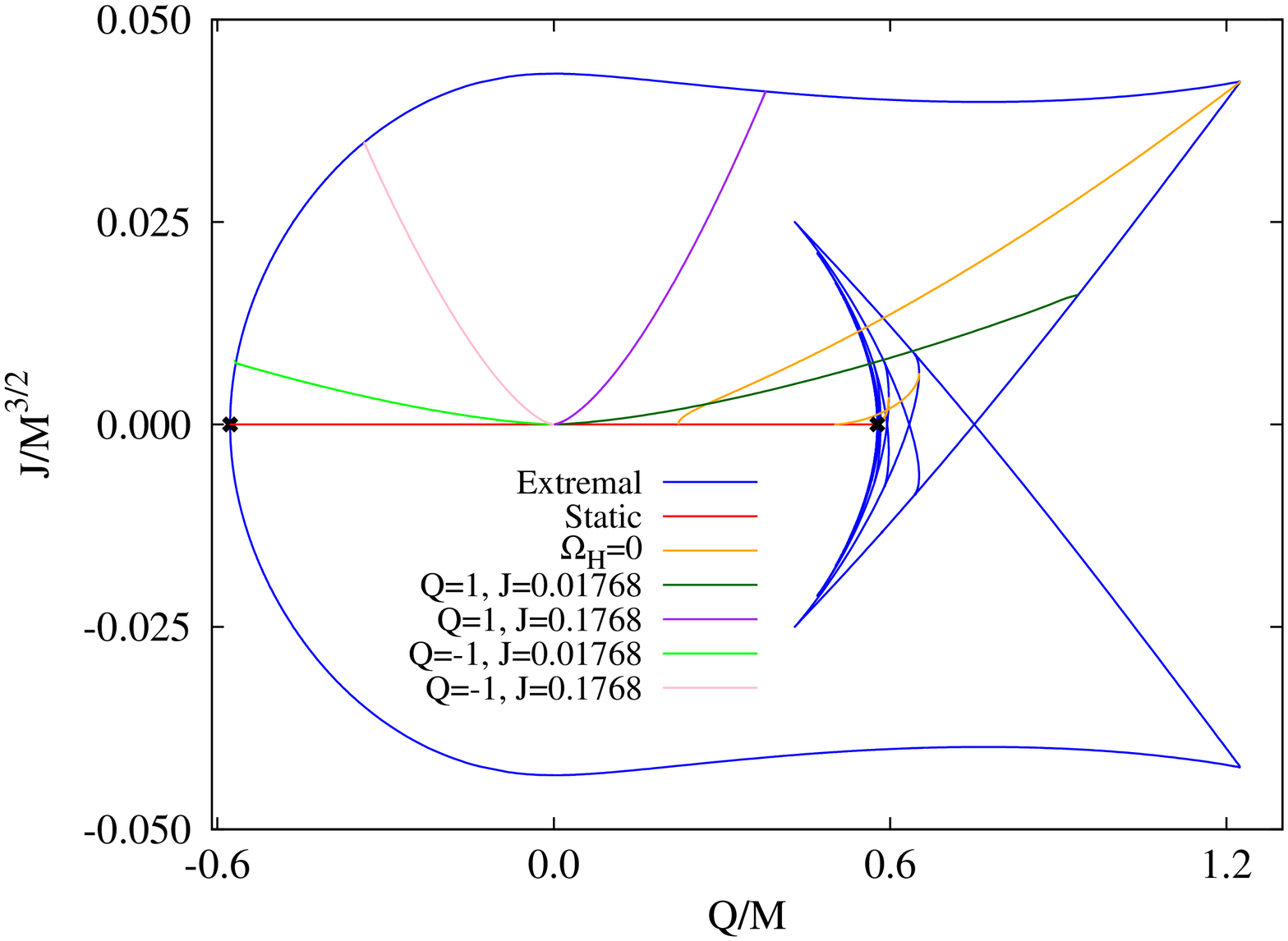}
\label{plot_q_j_lambda_5_extremal_z2b_long_paper_new_Om0.ps}
}
}
\vspace{-0.5cm}
\mbox{\hspace{-1.5cm}
\subfigure[][]{
\includegraphics[height=.35\textheight, angle =270]{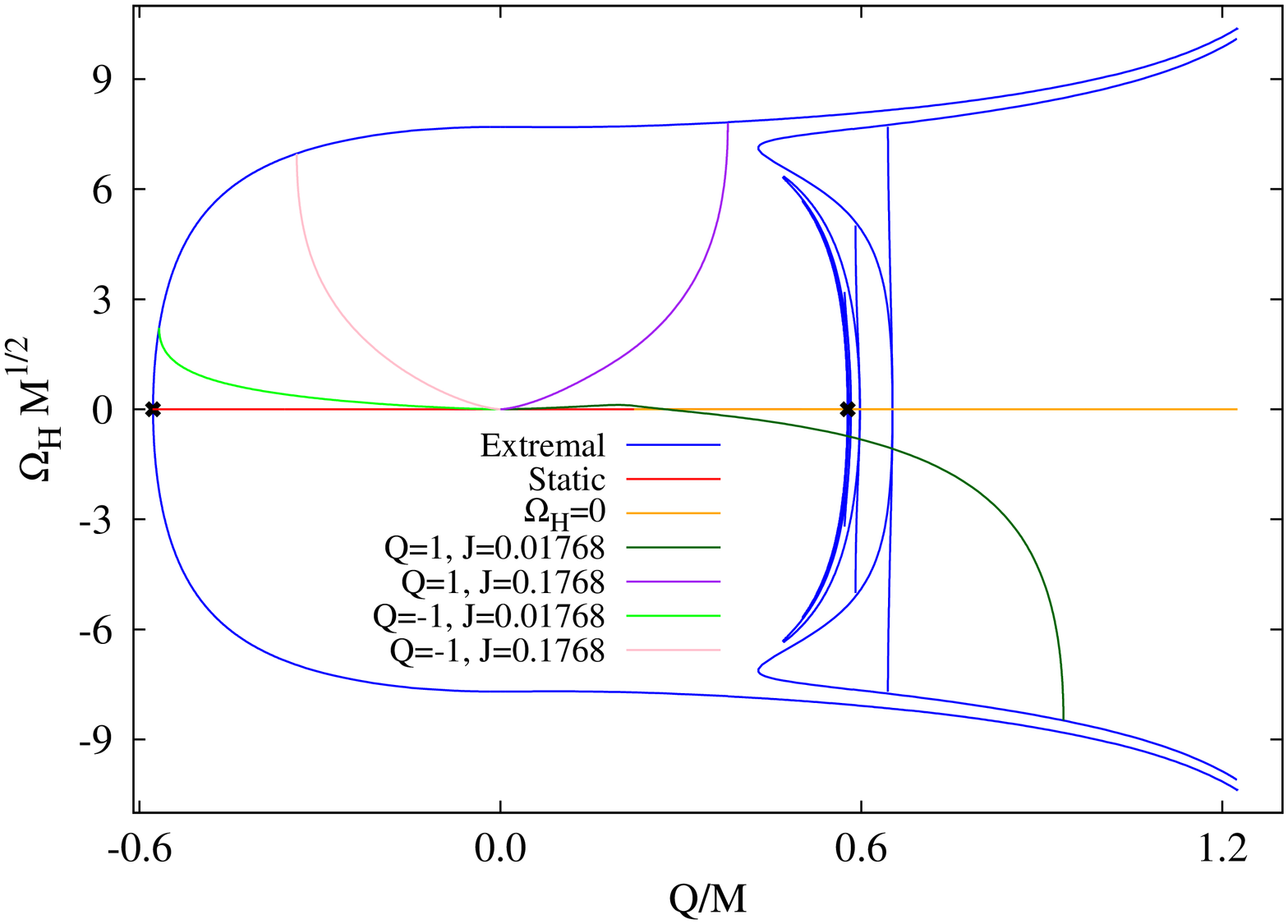}
\label{plot_q_omh_lambda_5_extremal_z2b_long_paper_new_Om0.ps}
}
\subfigure[][]{\hspace{-0.5cm}
\includegraphics[height=.35\textheight, angle =270]{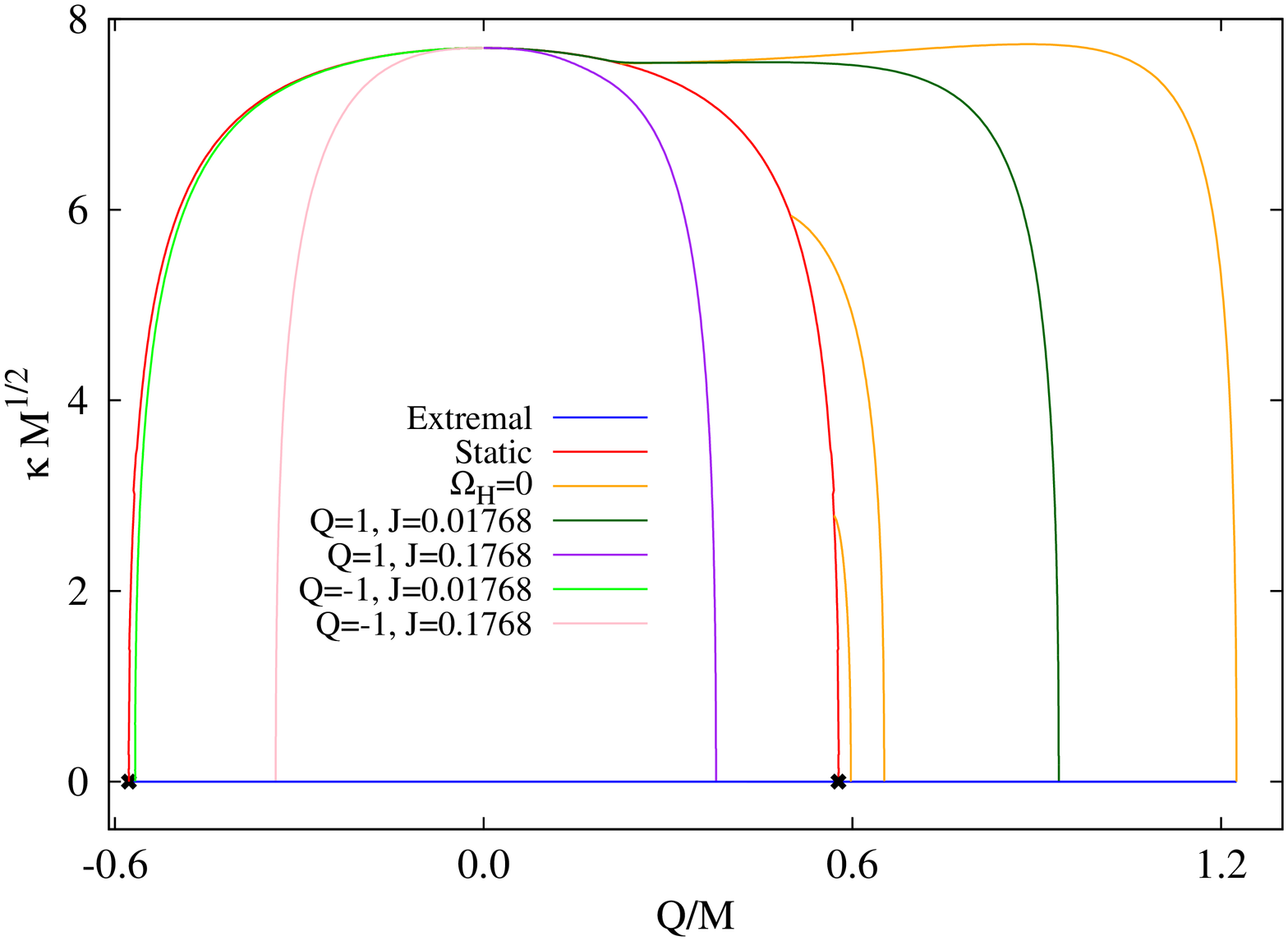}
\label{plot_q_kappa_lambda_5_extremal_z2b_long_paper_new_Om0.ps}
}
}
\end{center}
\caption{Global and horizon properties:
The scaled angular momentum $j = J/M^{3/2}$
of extremal solutions including  
the bifurcation points and cusps 
versus the scaled charge $q=|Q|/M$ (a).
The scaled angular momentum $j = J/M^{3/2}$
of extremal and selected sets of non-extremal solutions (b).
The scaled horizon angular velocity $\Omega_{\rm H} M^{1/2}$ (c),
and 
the scaled surface gravity $\kappa M^{1/2}$ (d)
for the same sets of extremal and non-extremal solutions.
(CS coupling $\lambda=5$)
\label{fig10}
}
\end{figure}

Finally, we would like to address the domain of existence of global 
EMCS black hole solutions. Here we would like to first
consider the global charges and, in particular, the angular momentum
versus the charge for fixed mass.
Typically, the boundary of the domain of existence
of black hole solutions is then formed by extremal solutions,
where the non-extremal solutions reside inside this boundary.
Therefore we exhibit in Fig.~\ref{plot_q_j_lambda_5_extremal_z2a_long_paper.ps}
the scaled angular momentum $j=J/M^{3/2}$ versus
the scaled charge $q=Q/M$ for extremal solutions,
choosing again CS coupling $\lambda=5$.
(A similar picture is found for other values of $\lambda>2$.) 
The figure includes the first few
bifurcation points, cusps and non-static $J=0$ solutions.
This allows us to connect the locations of
the solutions in the domain of existence with their
locations in the previous figures. 

Fig.~\ref{plot_q_j_lambda_5_extremal_z2a_long_paper.ps} then shows, that
when the scaled angular momentum is considered versus the scaled charge,
the domain of existence is indeed delimited by 
branches of extremal solutions, which constitute its boundary. 
Clearly, the domain of existence is different for positive and negative charge solutions,
since for spinning solutions the CS term breaks the charge reversal symmetry, $Q\to -Q$.
The extremal solutions with negative charge form the smooth $Q<0$ boundary.
This contains the extremal static RN solution, when the angular momentum vanishes.
The extremal solutions with positive charge, on the other hand,
reflect the complicated branch structure discussed above.

For positive charge only a subset of the extremal solutions forms the
$Q>0$ boundary. This subset of solutions starts at the
extremal Myers-Perry ($q=0$) solutions, passes the cusps
with vanishing horizon area, and ends at the first ($n=1$) non-static $J=0$ solution. 
Thus in contrast to the $Q<0$ boundary, 
the $q>0$ boundary is not smooth, since it
contains the pair of symmetric $J \to -J$ cusps, which reside
at the maximal value $q_{\rm max}$ of the scaled charge $q$. 
The solutions at the cusps correspond to the singular solutions 
with vanishing horizon area in Fig.~\ref{fig4}.
Moreover, this maximal value of $q$ is larger than the value 
of $q$ of the corresponding extremal RN solution. 

Even more surprising, however, is the fact, that
there are extremal solutions in the interior of the domain of existence. 
In particular, the $Q>0$ extremal RN solution is no longer part of the boundary.
Such a feature does not occur in theories like pure Einstein-Maxwell theory, 
or Einstein-Maxwell-dilaton theory, and neither for EMCS theory for
sufficiently low values of the CS coupling constant.
This intriguing fact produces a peculiar type of uniqueness violation: 
there exist both extremal and non-extremal solutions,
which possess the same global charges, $M$, $J$, and $Q$.

For more clarity and later reference, we replicate the
demonstration of the domain of existence 
in Fig.~\ref{plot_q_j_lambda_5_extremal_z2b_long_paper_new_Om0.ps},
now without the inset, the bifurcations and the cusps, but instead
with the inclusion of a number of selected sets of non-extremal solutions.
These sets contain, in particular, the branches of non-static solutions 
with non-rotating horizon, $\Omega_{\rm H}=0$.
Clearly, all solutions lie within the domain of existence,
while the crossings of lines exhibit the non-uniqueness of the solutions.

Let us next consider the horizon properties of the solutions.
In Fig.~\ref{plot_q_omh_lambda_5_extremal_z2b_long_paper_new_Om0.ps}
we exhibit the scaled horizon angular velocity $\Omega_{\rm H} M^{1/2}$
for the extremal solutions and the same set of non-extremal solutions.
Most of the boundary for the scaled horizon angular velocity $\Omega_{\rm H} M^{1/2}$
is formed by a subset of the extremal black holes, only
the righthand boundary, $i.e.$, the large $q$ limit,
is given by the maximum value of $q$, $q_{\rm max}$.
All other extremal and all non-extremal solutions reside within these boundaries.

We exhibit the scaled surface gravity $\kappa M^{1/2}$ 
in Fig.~\ref{plot_q_kappa_lambda_5_extremal_z2b_long_paper_new_Om0.ps}.
Clearly, the extremal black holes form the lower part of the boundary,
since they all possess $\kappa=0$.
For the $Q<0$ solutions, the static black holes form the upper part
of the boundary, as is often the case.
However, for the $Q<0$ solutions, the static solutions represent only
the first part of the upper part of the boundary.
Then another set of solutions adopts the role
of forming the upper part of the boundary. 
Interestingly, this set corresponds precisely to
the first set of non-static $\Omega_{\rm H}=0$ solutions,
which start at a certain static RN solution and reach all the way to
the singular cusp.

Let us next address the scaled area $a_{\rm H} = A_{\rm H}/M^{3/2}$
.
 For $Q<0$, the extremal and the static solutions form again together the boundary.
For $Q>0$, the upper boundary part is also formed 
the first part of the set of static solutions, 
until the first set of non-static $\Omega_{\rm H}=0$ solutions emerges from 
the static solutions, which then form the remaining upper part of the boundary.

Finally, we address
the gyromagnetic ratio $g$ of the same sets of black hole solutions.
Again, for $Q<0$, the gyromagnetic ratio is bounded by
the static and extremal solutions. For $Q>0$, however, 
the gyromagnetic ratio becomes unbounded, as was noted before 
\cite{Kunz:2005ei,Kunz:2006yp}.

\section{Conclusions}

Here we have discussed global and near-horizon solutions of EMCS theory in five
dimensions, focussing on solutions with equal magnitude angular momenta,
to enhance the symmetry of the solutions, making the analytical and numerical
analysis much more tractable while at the same time revealing already numerous
intriguing features of the solutions.

Since the CS term breaks the charge reversal invariance, the families of solutions
for negative charge and positive charge no longer agree, when the CS term
contributes, $i.e.$, for rotating solutions.
Choosing the CS coupling constant $\lambda$ to be positive, then the negative
charge solutions do not represent any particular peculiarities.
The positive charge solutions, however, develop a number of very interesting 
properties, as the CS coupling constant $\lambda$ first reaches 
the supergravity value $\lambda_{\rm SG}$ and then exceeds it.

At $\lambda_{\rm SG}$ the solutions are known analytically \cite{Chong:2005hr},
with the BMPV solutions \cite{Breckenridge:1996is} forming a special case,
which represent extremal non-static $\Omega_{\rm H}=0$ solutions,
ending in a singular $A_{\rm H}=0$ solution. 
For $\lambda \ne \lambda_{\rm SG}$ the global solutions are known only 
perturbatively \cite{Aliev:2008bh,Allahverdizadeh:2010xx} or numerically
\cite{Kunz:2005ei,Kunz:2006yp,Blazquez-Salcedo:2013muz}, obtained by solving
the corresponding set of ordinary differential equations, subject to
a set of appropriate boundary conditions.
However, the near-horizon solutions can be obtained in a simpler way,
by making use of an appropriately modified near-horizon formalism,
where the modification is caused by the CS term.

\begin{figure}[h!]
\begin{center}
\mbox{\hspace{-1.5cm}
\subfigure[][]{
\includegraphics[height=.35\textheight, angle =270]{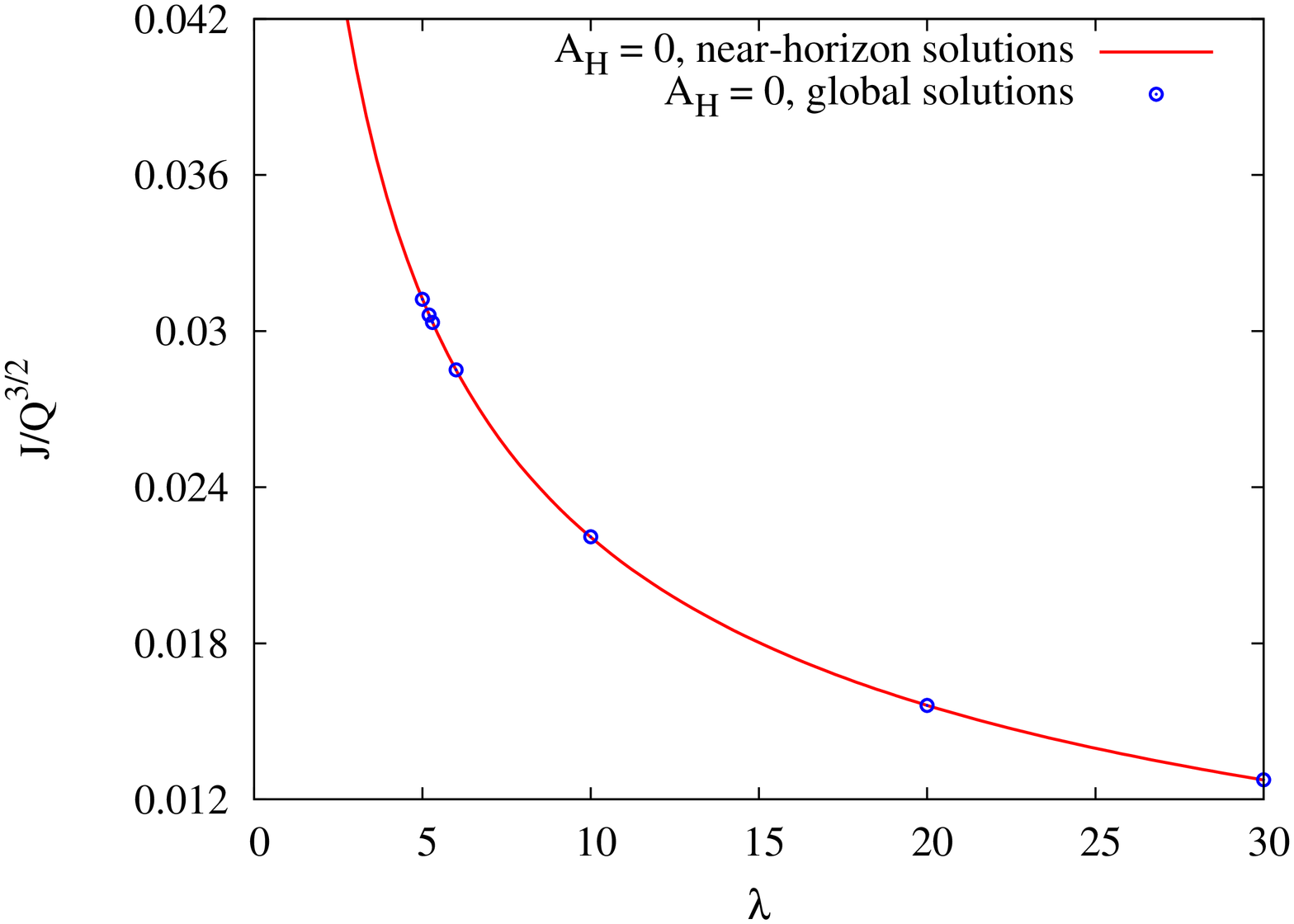}
\label{plot_q_j_lambda_MP_to_singular_branch_NH_formula.ps}
}
\subfigure[][]{\hspace{-0.5cm}
\includegraphics[height=.35\textheight, angle =270]{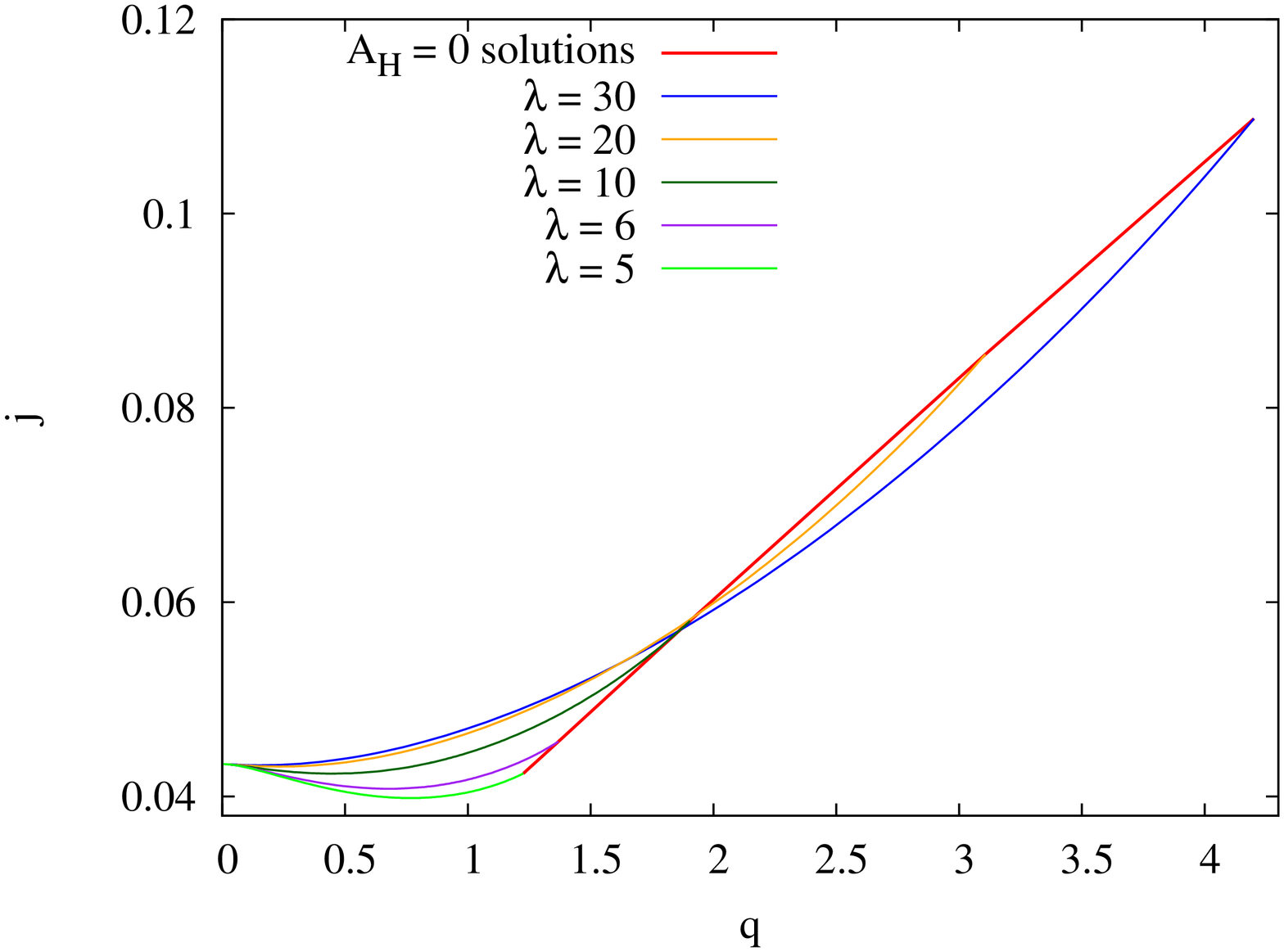}
\label{plot_q_j_lambda_MP_to_singular_branch_2.ps}
}
}
\end{center}
\caption{\small
Singular $A_{\rm H}=0$ solutions:
The scaled angular momentum $J/Q^{3/2}$
versus the CS coupling constant
for near-horizon and global solutions (a);
the scaled angular momentum $j = J/M^{3/2}$
versus the scaled charge $q=|Q|/M$ 
for global solutions (between the MP solution
and the $A_{\rm H}=0$ cusp) and several values of $\lambda$.
}
\label{fig11}
\end{figure}

The presence of a CS term always implies the occurrence of (a set of two
degenerate $J \to -J$ symmetric)
singular $A_{\rm H}=0$ solutions. But only for $\lambda \ge \lambda_{\rm SG}$
these singular solutions appear prominent in the domain of existence,
residing at a cusp. We exhibit the $\lambda$-dependence of these solutions
in Fig.~\ref{fig11}. 
The cusp points have been obtained both for the near-horizon
and the global solutions, as demonstrated in 
Fig.~\ref{plot_q_j_lambda_MP_to_singular_branch_NH_formula.ps}.
When the cusp points are considered for the scaled angular momentum
and the scaled charge, $i.e.$, in the form $j(q)$,
we note, that they form an almost straight line,
as depicted in Fig.~\ref{plot_q_j_lambda_MP_to_singular_branch_2.ps}.

Here we have focussed our considerations on 
values of the CS coupling constant $\lambda > 2 \lambda_{\rm SG}$,
since above this limit a new type of extremal solutions is present:
extremal non-static $J=0$ solutions.
These solutions come as degenerate pairs, symmetric with respect to
$J \to -J$, where the lowest solution resides in the boundary of existence
of the black hole solutions. The solutions form a presumably infinite sequence,
that can be labeled by an integer $n$. This integer is associated
with the number of nodes of the metric function $\omega$ and 
the gauge field function $a_\varphi$.
To our knowledge, this is the first example of
black holes with Abelian fields which form excited states,
which are reminiscent of the radial excitations of atoms, 
of sphalerons, or of hairy black holes.

With increasing node number the mass of these extremal non-static $J=0$ black holes
converges to the mass of the extremal static RN black hole,
while their horizon angular velocity tends to zero
along with their magnetic moment.
Their area, however, retains the same constant value,
which is different from the static value.
We demonstrate this pointwise convergence of the solutions in the Appendix.

The non-static $J=0$ solutions are located symmetrically within an intriguing
pattern of branches of extremal solutions.
Most of these branches of extremal solutions 
reside within the domain of existence of EMCS black holes,
when the scaled angular momentum is considered versus the scaled charged.
This is rather unusual, since the extremal solutions typically form the 
boundary of the domain of existence in this case.
It has the interesting consequence, that there is non-uniqueness
between extremal and non-extremal black holes.
Furthermore, there is non-uniqueness between extremal black holes,
which reside at the bifurcation points of the branches,
while non-uniqueness between non-extremal black holes was reported already
earlier \cite{Kunz:2005ei}.

The comparison between near-horizon solutions and global solutions has
also led to surprises, namely we have seen, that
{\it a given near horizon solution
can correspond to}
i) {\it more than one global solution}, {\it possibly even an infinite set},
ii)
{\it  precisely one global solution}, or
iii)
{\it no global solution at all}.
 Thus our findings show that the intuition based on known exact solutions
cannot be safely applied in the general case.
Moreover, the results of a near-horizon analysis of extremal solutions cannot be
always safely extrapolated to the global case, 
as assumed sometimes in the literature.

Similar black holes can be obtained in EMCS theory in higher odd dimensions, when all the angular momenta have the same magnitude \cite{Kunz:2006yp}. In 
particular, our preliminary results show that for $D=7, 9$ this sequence of non-static $J=0$ solutions is also found, and the solutions can
be characterized by the number of nodes of the corresponding metric and gauge field functions.

We conjecture that extremal black holes with similar properties
may also exist in other theories, in particular, in an Einstein-Maxwell-dilaton
theory in four dimensions \cite{Kleihaus:2003df}.
In this case, however, a set of partial differential equations must be solved,
making the corresponding analysis much more involved.

However, our next step will be the inclusion of a cosmological constant for these
EMCS black holes. When the new intriguing phenomena observed for the
asymptotically flat solutions survive in the presence of 
a negative cosmological constant, 
the AdS/CFT correspondence may imply interesting consequences for 
the associated four-dimensional field theories, living on the boundary.

\medskip
{\bf Acknowledgement}
\\
We gratefully acknowledge support by 
the DFG Research Training Group 1620 ``Models of Gravity''
and by the Spanish Ministerio de Ciencia e Innovacion, 
research project FIS2011-28013.
The work of 
E.R. is supported by the FCT-IF programme and the
CIDMA strategic project  UID/MAT/04106/2013.

\section{Appendix}

In this appendix we briefly illustrate the full set of functions
and the dependence of the nodes for the sequence of non-static $J=0$ extremal 
solutions
and compare with the corresponding RN functions.

\begin{figure}[h!]
\begin{center}
\mbox{\hspace{-1.5cm}
\subfigure[][]{
\includegraphics[height=.35\textheight, angle =270]{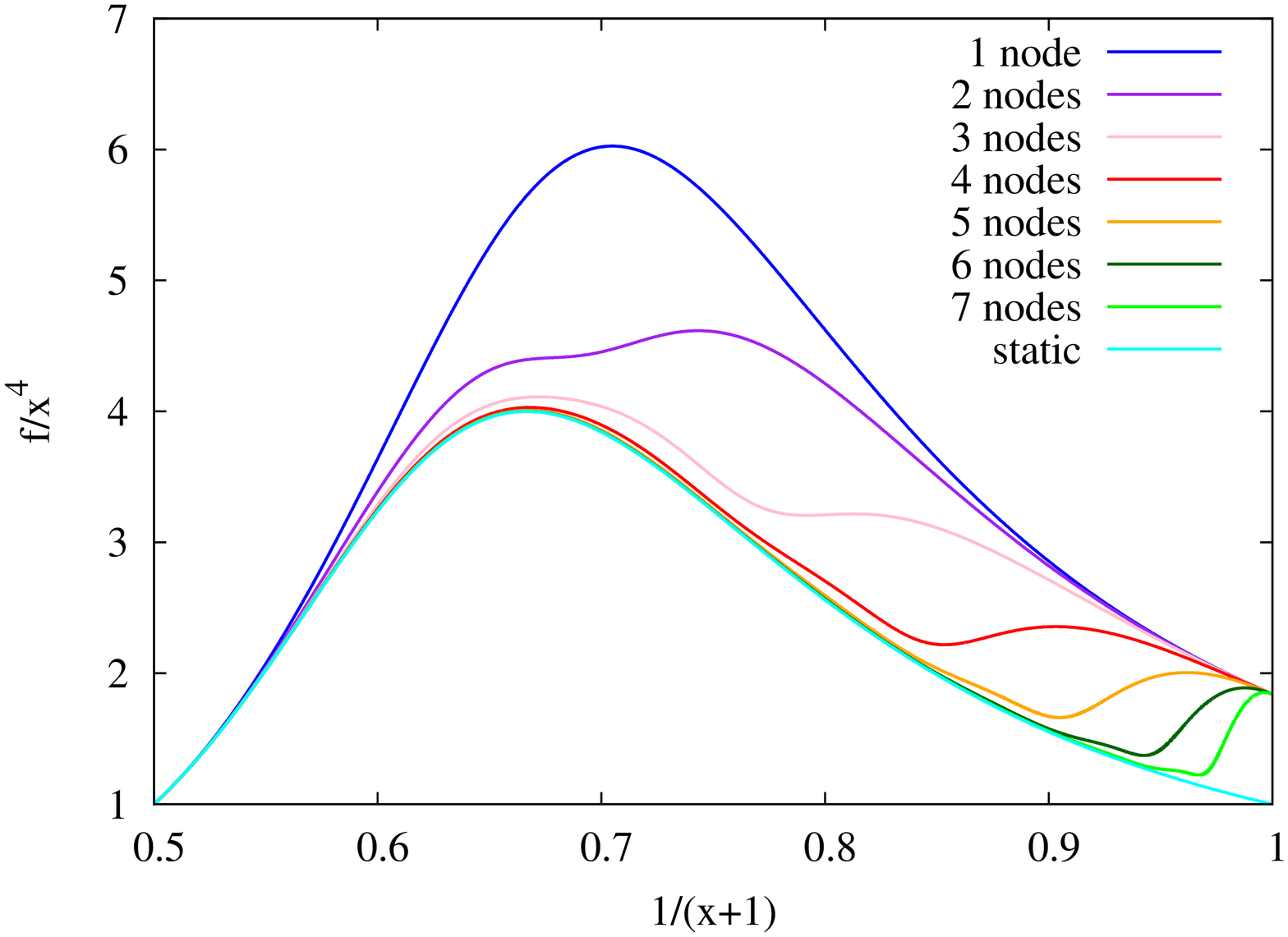}
\label{plot_x_f_lambda_5_J=0.ps}
}
\subfigure[][]{\hspace{-0.5cm}
\includegraphics[height=.35\textheight, angle =270]{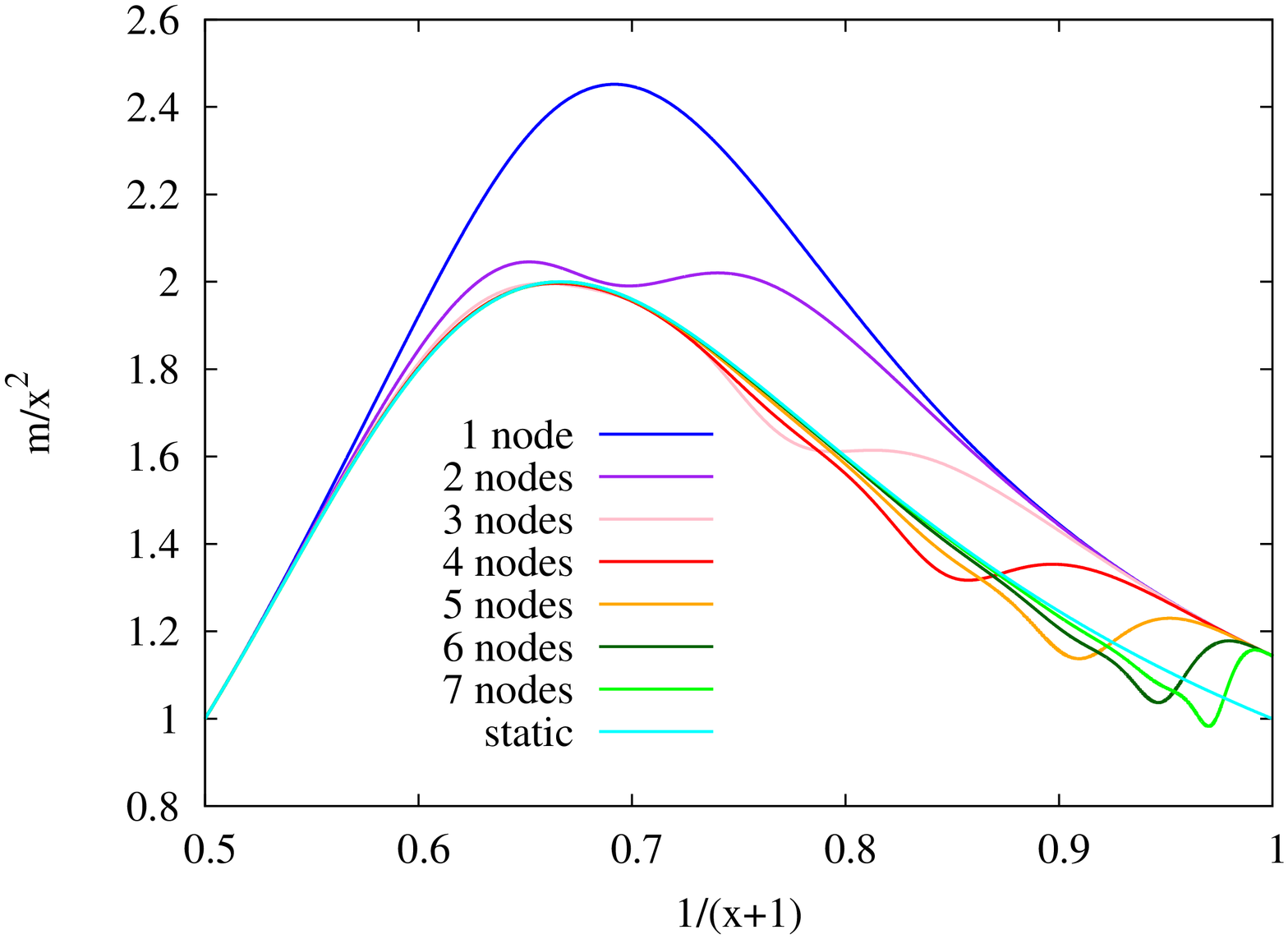}
\label{plot_x_m_lambda_5_J=0.ps}
}
}
\vspace{-0.5cm}
\mbox{\hspace{-1.5cm}
\subfigure[][]{
\includegraphics[height=.35\textheight, angle =270]{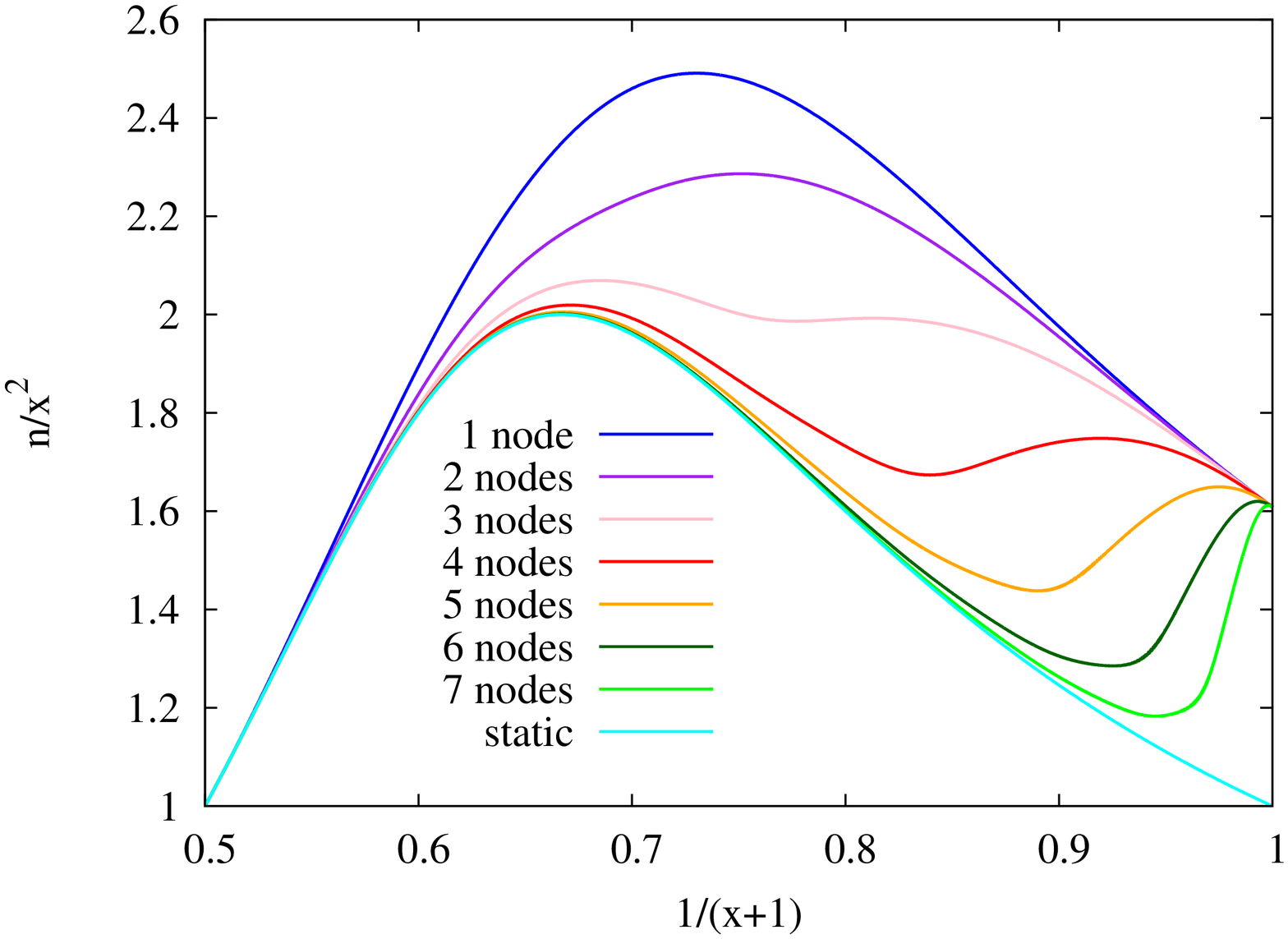}
\label{plot_x_n_lambda_5_J=0.ps}
}
\subfigure[][]{\hspace{-0.5cm}
\includegraphics[height=.35\textheight, angle =270]{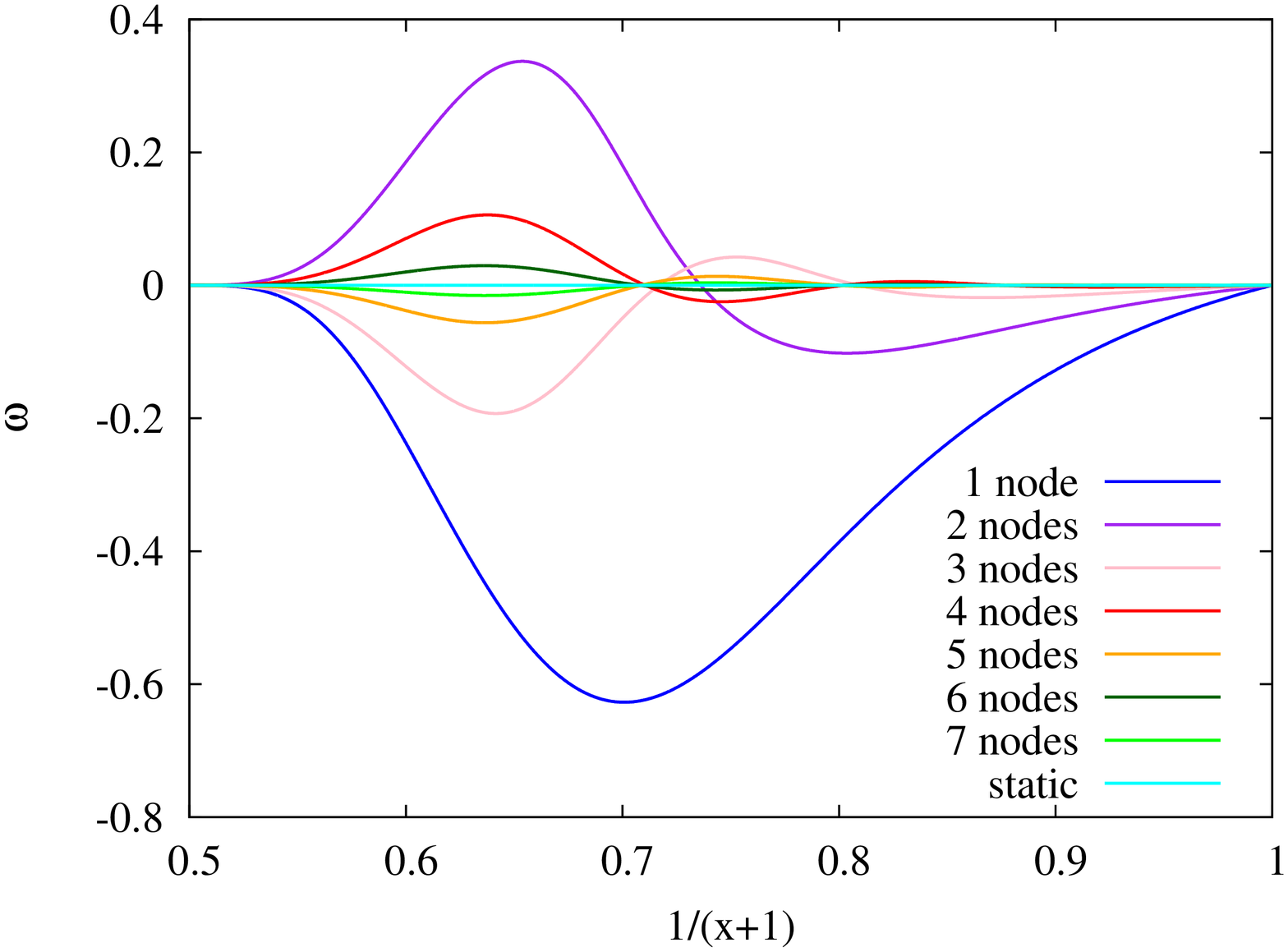}
\label{plot_x_omega_lambda_5_J=0.ps}
}
}
\vspace{-0.5cm}
\mbox{\hspace{-1.5cm}
\subfigure[][]{
\includegraphics[height=.35\textheight, angle =270]{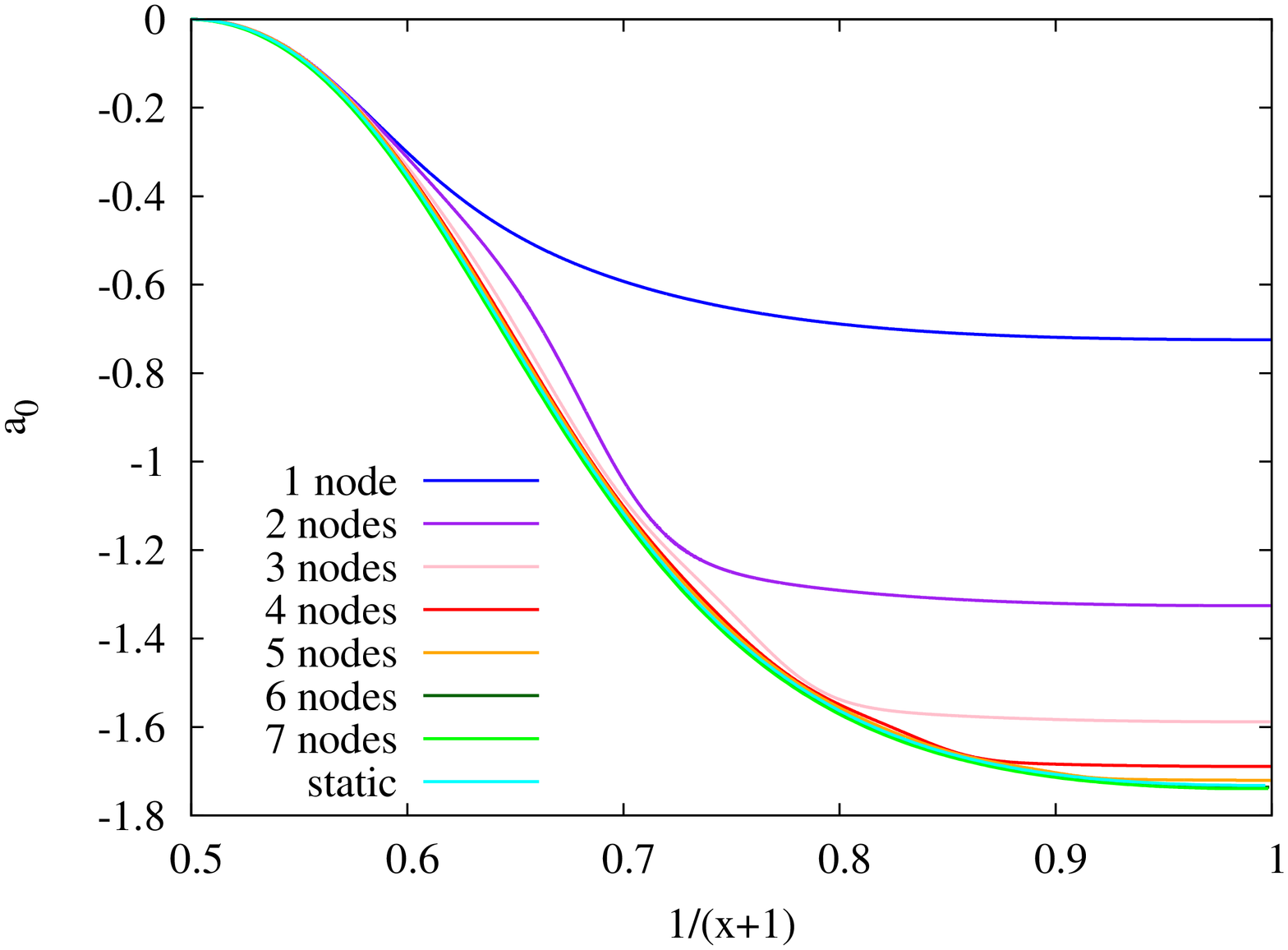}
\label{plot_x_a0_lambda_5_J=0.ps}
}
\subfigure[][]{\hspace{-0.5cm}
\includegraphics[height=.35\textheight, angle =270]{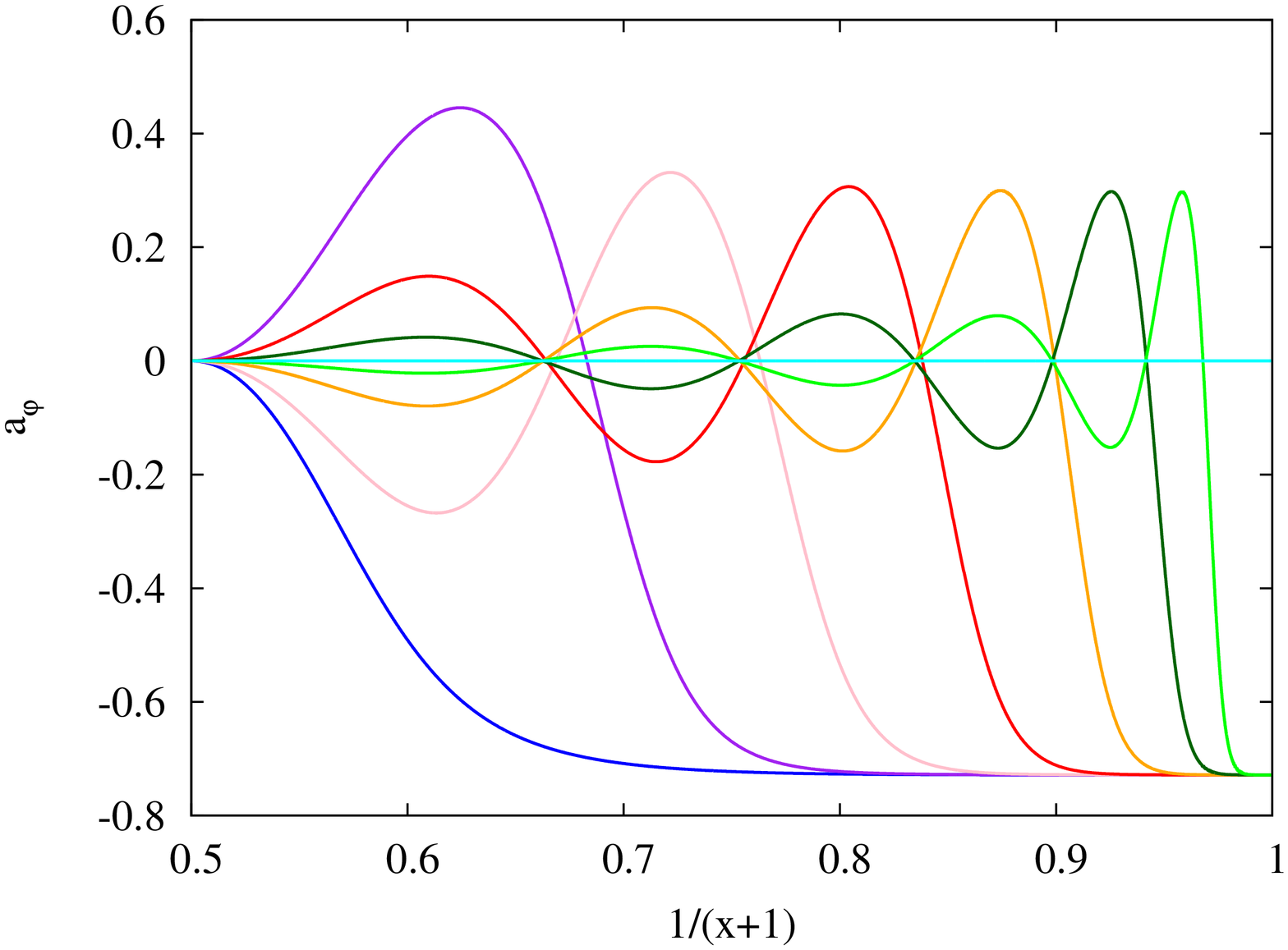}
\label{plot_x_ak_lambda_5_J=0.ps}
}
}
\end{center}
\caption{\small
Global $J=0$ solutions:
The metric function $f$ (a), $m$ (b), $n$ (c), $\omega$ (d)
and the gauge field functions $a_0$ (e) and $a_\vphi$ (f)
versus the radial coordinate
for the lowest radial excitations $n=1,...,7$.
For comparison, the corresponding functions for the extremal RN solution
are also shown.
(Charge $Q=1$ and CS coupling $\lambda=5$).
}
\label{fig12}
\end{figure}

We present in Fig.~\ref{fig12} the radial dependence of the functions of the
fundamental $J=0$ solution and the first few excitations.
We note that the functions $f$, $m$, $n$ and $a_0$ exhibit a fast convergence
towards a limiting solution, which is represented by the static extremal RN
solutions. However, this convergence is only local. At the horizon, $x=0$,
the functions do not tend to the RN values in most cases.
In the limit $n \to \infty$ the functions will then exhibit a jump.
We notice this also for the function $a_\varphi$.
Such a discontinuity is necessary, to guarantee that these solutions,
while approaching the corresponding RN solution, retain a 
horizon area which differs from the RN horizon area, 
but remains the same for all solutions of the sequence.

\end{document}